\documentclass[usenatbib]{mn2e}

\input psfig.tex

\usepackage{graphicx}
\usepackage{natbib}
\usepackage{array}
\bibpunct{(}{)}{;}{a}{}{,}

\usepackage{rotating}    

\input psfig.tex
\usepackage{latexsym}
\usepackage{natbib}
\usepackage{amssymb}
\usepackage{amsmath}

\usepackage{graphicx}
\usepackage{graphics}
\usepackage{fancyhdr}
\usepackage{morefloats}

\title[Early type galaxies formation and evolution]{Formation and Evolution of Early-Type Galaxies. III \\
 Dependence of the star formation history on the total mass and initial over-density}

\author[E. Merlin et al.]{E. Merlin$^{1}$\thanks{E-mail: emiliano.merlin@unipd.it},
		C. Chiosi$^{1}$, L. Piovan$^{1}$, T. Grassi$^{1}$, U. Buonomo$^{1}$ and F. La Barbera$^{2}$ \\
$^{1}$Department of Astronomy, University of Padova, Vicolo dell'Osservatorio 3, 35122 Padova, Italy  \\
$^{2}$INAF-Osservatorio Astronomico di Capodimonte, Salita Moiariello 16, Napoli, Italy }

\begin{document}

\date{Accepted ... Received ...}

\pagerange{\pageref{firstpage}--\pageref{lastpage}} \pubyear{2012}

\maketitle

\label{firstpage}


\begin{abstract}

We investigate the influence of the initial over-densities and masses of proto-galaxies on their subsequent evolution (the star formation history in particular), to understand whether these key parameters are sufficient to account for the varied properties of the galactic populations. By means of fully hydrodynamical N-body simulations performed with the code \textsc{EvoL},  we produce twelve self-similar models of early-type galaxies of different initial masses and over-densities, and follow their evolution from the early epochs (detachment  from the linear regime and Hubble flow at $z \geq 20$) down to the stage when mass assembly is complete i.e. $z\leq1$ (in some cases the models are calculated up to $z=0$. The simulations include radiative cooling, star formation, stellar energy feedback, re-ionizing photo-heating background, and chemical enrichment of the interstellar medium; we do not consider the possible presence of Active Nuclei. We find a strong correlation between the initial properties of the proto-haloes and their subsequent star formation histories. Massive ($M_{tot}\simeq 10^{13} M_{\odot}$) haloes experience a single, intense burst of star formation (with rates $\geq 10^3 M_{\odot}$/yr) at early epochs, consistently with observations, with  less pronounced dependence on the initial over-density; intermediate mass ($M_{tot}\simeq 10^{11} M_{\odot}$) haloes have histories that strongly depend on their initial over-density, whereas low mass haloes ($M_{tot}\simeq 10^{9} M_{\odot}$)  always have erratic, bursting like star forming histories, due to the ``galactic breathing'' phenomenon. The model galaxies have morphological, structural, and chemical   properties resembling  those of real galaxies, even though some disagreement still occurs, likely a consequence of some numerical choices.  We conclude that total mass and initial over-density drive  the  star formation histories of early type galaxies. The model galaxies belong to the so-called quasi-monolithic (or early hierarchical) scenario in the sense the aggregation of lumps of dark and baryonic matter is completed very early on in their history. In this picture, \textit{nature} seems to play the dominant  role, whereas \textit{nurture} has a secondary importance.
\end{abstract}




\section{Introduction} \label{intro}

Understanding how early-type galaxies (ETGs) formed and evolved along the Hubble time is still a formidable challenge. According to the current view of the subject, galaxies probably began to form at $z \sim 20 - 50$, when Dark Matter (DM) haloes containing Baryonic Matter in cosmological proportions gave origin to the first sufficiently deep gravitational potential wells \citep{Tegmark1997,Gao2007a,Gao2007b}. However, it is not yet clear how the paradigm $\Lambda$-CDM cosmological model (which prescribes a hierarchical, bottom-top formation of structures, implying that
massive systems assemble their mass later than the smaller ones) could be reconciled with the observational evidence for large, and red, galaxies already in place at very high redshifts \citep[see e.g.][]{Marchesini2010, Harrison2011, Mortlock2011}. The problem is twofold: on one hand, it must be explained how and when star formation is quenched in both massive and small haloes, which is necessary to reconcile the theoretical prediction with the observed
galaxy mass function \citep[see e.g.][]{Bundy2006}; on the other hand, it should be also clarified how such massive systems can form at very high redshifts. Semi-analytical models of hierarchical formation
\citep[e.g., ][]{DeLucia2006,DeLucia2007,Almeida2007,Gonzalez2009,Parry2009,DeLucia2011} have succeeded in explaining some of the observational features of ETGs, but fail in others. In particular they overestimate the number of faint objects and underpredict the sizes of bright ETGs. In addition to this, the hierarchical formation finds it difficult to explain the trend shown by the  so-called $\alpha$-enhancement at varying the mass of the galaxy. While massive ETGs are enhanced in $\alpha$-element, the low mass ones are not. This requires that the history of star formation varies with the galaxy mass in a way that can be hardly met by the hierarchical scheme. However, since the topic has been examined in detail by \citep{Tantalo2002} and no new arguments have been made available we will not discuss it in details (but see Sect. \ref{populations}).

All the  above issues are   essentially a reformulation of the long-standing dichotomy between the so-called \textit{hierarchical} and \textit{monolithic} models of galaxy formation, as described respectively e.g. by \citet{Eggen1962} and by \citet{WhiteRees1978}, possibly including some more sophisticated features \citep[e.g. the \textit{dry merger} scenario, as proposed by][]{Bell2004}. After the Cold Dark Matter cosmological model has got worldwide acceptance, the hierarchical model appeared to be the most convincing scenario of galaxy formation. In recent times, however, the situation has become by far more intrigued than in the past. Evidences claiming for a new scenario
very close indeed to the classical monolithic one have long been known both observationally and theoretically \citep[see e.g.][]{Chiosi2002}, but recently they have become compelling \citep[e.g.][]{Thomas2011,Gobat2011}.

\citet{Merlin2006, Merlin2007} suggested that a series of primordial mergers of small stellar subunits (in a sort of early hierarchical galaxy formation mechanism), taking place at $z \geq 2$, could lead to the formation of massive objects in the early Universe, resembling the observed systems as far as many different aspects (morphology, density profiles, metallicity...) are concerned. This picture finds support in numerical simulations of  the very first
stellar populations in the early universe, such as those by \citet{Bromm2002} and \citet{Yoshida2003}, who  found that the first virialized objects might form at $z \sim 30$ with a typical mass of $10^6 M_{\odot}$. In this scheme, a first generation of stars in clumps inside primordial haloes of DM would first enrich the medium in metals, and also act as the building blocks of larger systems with masses up to $10^{12} M_{\odot}$ via gravitational clustering. The mode of star formation and the efficiency of it is likely driven by the mass and mean density of the proto-galaxy as long ago pointed out  by \citet[][]{Chiosi2002} and more recently argued on observational basis by \citet[][]{Harrison_etal_2011}.

A fundamental piece of the puzzle is the role played by energy feedback during the formation of the galactic systems. Although it has long been known  \citep[see e.g. ][]{McKee1977} that supernova explosions can inject large amounts of kinetic and thermal energy into the galactic Interstellar Medium (ISM), only more recently the importance of this phenomenon as a source of pressure support and large scale turbulent motions is being fully recognized. The effects
of the energy injection are still under investigation, since two contrasting consequences are expected to follow - i.e. the quenching of the star formation activity due to the increasing of the local gas pressure (\textit{negative feedback}), and  the  indirect enhancement of the stellar activity  due to the formation of layers of cold and dense gas within the turbulent, shocked ISM (\textit{positive feedback}). However, on larger scales bubbles and super-bubbles of hot and thin material are expected to develop due to the action of energy injection, expanding in the ISM and giving rise to galactic winds. Detailed simulations by \citet{Ceverino2009} showed that the correct modeling of the energy injection from supernova explosions is sufficient to eject a large fraction of the gaseous content from a galactic disc. How this phenomenon is expected to influence the evolution of massive spheroids is still to be clarified. On the other hand, it is widely accepted that the energy release from Active Nuclei (AGN) should be a fundamental ingredient
for quenching the star formation activity in such systems. However, in this study we show  how stellar feedback may be sufficient to achieve the goal, provided it is modeled in an efficient way.

Recently, many numerical studies have focused on the formation of individual galaxies instead of large scale simulations. Among the most recent ones, \citet{Sales2010} produced different galaxy models with different feedback prescriptions; \citet{Sommer-Larsen2010} simulated a cluster and studied its evolution and the galactic populations; \citet{Stinson2010} used a multi-resolution technique to study the formation of galaxies within high-mass haloes; \citet{Croft2009} examined a  set of models extrapolated from a large scale cosmological simulation; \citet{Naab2007} presented three models produced \textit{without} feedback of any kind; \citet{Cox2006} investigated the merger of  spirals; \citet[][]{Kobayashi2004,Kobayashi2005} calculated a very large number of low resolution chemo-dynamical models of elliptical galaxies; \citet{Kawata1999} presented models of elliptical galaxies formed out of cosmological perturbations; finally, \citet[][]{Chiosi2002} presented the first chemo-dynamical N-body Tree-SPH (NB-TSPH) simulations of galaxy formation based on the monolithic scheme, including star formation, cooling, energy feedback, chemical enrichment and galactic winds at varying the initial total mass and density of the proto-galaxy, and brought into evidence that a continuous transition of the star formation mode from a single initial episode down to a a recurring series of star forming events is possible at varying these two parameters.
Each of these studies addressed to a particular aspect of galaxy formation.
Although the standard hierarchical scheme is fully coherent with the $\Lambda$-CDM cosmology, many observational hints seem to indicate that the revised monolithic scheme in which very early mergers of subunits made of gas and stars are taking place yields theoretical models that are able to explain a wide range of observational data for ETGs. Therefore, it is worth exploring in more details the possibilities offered by this scheme as a viable and perhaps complementary view of the galaxy formation problem. It might be that in the $\Lambda$-CDM context while DM haloes aggregate and merge building up ever increasing large scale structures, very early on baryons collapse in the gravitational potential wells of DM mimicking the monolithic like mode.

\textbf{Aims of the present study}. We intend to explore the possibilities of the monolithic (or early hierarchical) mode of galaxy formation we have just outlined and simultaneously to bridge a gap between the results presented by
\citet[][]{Chiosi2002} and those by \citet{Merlin2006,Merlin2007}. 
In the first study, the initial conditions for the proto-galaxy were not derived from self consistent cosmological simulations, but were created \textit{ad-hoc} to model a spherical lump of DM + BM (in the right proportions depending on the adopted cosmological background) subject to collapse: the DM particles were distributed following the \citet{Navarro_etal_1996} universal profile, with velocities derived from the velocity dispersion $\sigma(r)$ for a spherical, isotropic, and collision-less system with the adopted density profile \citep{Binney_Tremaine1987}; while the BM particles (in form of gas at the beginning) were distributed homogeneously inside the DM halo with zero velocity field, mimicking the infall of primordial gas into the potential well of DM \citep{WhiteRees1978}. From this stage, star formation in the proto-galaxies was allowed to occur. Finally, supposing that at each redshift the over-density of proto-galaxies could vary within a suitable interval, many models were calculated at varying the total mass and initial over-density.

In the two studies by \citet{Merlin2006,Merlin2007}, the initial conditions for DM and BM particles of the proto-galaxies were derived from large scale cosmological simulations, however adapted to our purposes. Since the procedure is much similar to the one adopted here, they will be jointly described below. However, in those papers no attempt was made to confirm the \citet[][]{Chiosi2002} scenario about the mass and/or initial density dependence of the star formation history of a galaxy.

With this target in mind, the NB-TSPH code \textsc{EvoL} was fully revised, re-written in parallel language and much improving the input physics. All details about the physical ingredients and the many tests we have performed to validate the code can be found in \citet{Merlin2009, Merlin2010} to whom the reader should refer. The present paper completes the analysis initiated with \citet[][]{Chiosi2002} and continued with \citet[][]{Merlin2006,Merlin2007}.
Here, we focus on the relation between the initial conditions of the host halo and the final properties of the galaxy making use of the parallel code \textsc{EvoL} and a consistent cosmological scenario to set the initial conditions. We investigate the process leading to the formation of isolated galaxies with different total masses and initial over-densities and cast light on whether the internal properties of a ``real'', cosmologically consistent galactic halo made of DM and BM are sufficient to obtain the variety of structural and physical properties which characterize the population of early-type galaxies and to confirm the predictions made by \citet{Chiosi2002}. Furthermore, we will examine whether the size predicted by the new models better agree with the observational data compared to the results obtained by the semi-analytical hierarchical models we have mentioned above.

To this aim, we simulate the formation and evolution of twelve isolated early-type galaxy models, from the initial stages of the non-linear evolution of the proto-galactic haloes at very high redshift, down to at least redshift $z=1$, and beyond in most cases. To avoid any misunderstanding, we remind the reader that the evolution of the density perturbations, once exited from the linear regime, toward to the stage of proto-galaxy and of this latter to the present, are followed by means of NB-TSPH simulations, the only tool fully entitled to this task during the non-linear regime.

The dichotomy between \textit{nature} (that is  the initial properties of a proto-galactic system set \textit{ab initio} by the primordial cosmological fluctuations) and \textit{nurture} (that is, the action of environmental factors) in the process of galaxy formation is fiercely discussed nowadays. In this context, we claim that the initial over-density of an halo should be generally considered part of its \textit{nature}, since it is essentially fixed by the cosmological perturbations from which the halo forms, rather than part of the \textit{nurture} processes,
as it is sometimes viewed. For example, belonging to a group or a cluster of galaxies (that virialize at a certain epoch) imply that the mean over-density of the region is high, and therefore it is likely that galactic haloes belonging to the region had, on average, higher initial over-densities than their counterparts forming in the field.
Of course, isolated \textit{but} strongly over-dense galactic haloes are possible, but they are less probable. If the initial cosmological perturbed density field is a linear superposition of independent waves, having the same power on all scales in scale-free models, then the highest peaks of the density field are expected in regions having, on average, larger over-densities, rather than being isolated. In this picture, the only ``real'' \textit{nurture} processes are those relative to the action of nearby objects in terms of feedbacks, ram pressure, gas stripping,
galactic mergers, etc. The argument we made here is that these processes, while being of great importance in some cases, are not those leading the basic mechanisms of galaxy formation and evolution.

By varying the initial halo over-density of our models (in the way we are going to describe below), we try to mimic
the action of the ``natural'' over-density. In some way, one could consider our ``dense'' models as systems belonging to generally over-dense regions ultimately becoming groups or clusters of galaxies, and our ``less dense'' models as totally isolated (field) galaxies.

All the model galaxies are obtained from the same initial proto-halo, however suitably modified in order to systematically change its properties from model to model (see below). We also run a few more cases with different
prescriptions, to better explore the effects of some key physical quantities and assumptions.

The paper is subdivided as follows. In Sect. \ref{methods} we describe the numerical methods and the setting of the initial conditions. In Sect. \ref{results} we analyze the results of the simulations. In Sect. \ref{profiles} we discuss the mass density profiles of the stars we have obtained for the models. In Sect. \ref{MRR} we present a preliminary discussion of the Mass-Radius relationship for ETGs. Finally, in Sect. \ref{conclusions} we draw some final remarks and conclusions.

\section{Input Physics of the NB-TSPH  Model Galaxies} \label{methods}

\subsection{Setting of the initial conditions}

The ideal procedure to derive the initial structure of a proto-galaxy would be to start from large scale cosmological NB simulations of typically $\sim$500 Mpc on a side in the framework of a given cosmological model so that the appearance, growth and subsequent aggregation of perturbations of all scales can be suitably described, cfr. e.g. the Millennium Simulation described in \citet[][]{Springel2005b}  who studied the formation, evolution and clustering of DM haloes in the $\Lambda$-CDM cosmology. Of course these simulations require a huge number of particles: for instance, the Millennium run followed more than 10 billions particles. Considering the huge number of DM haloes (proto-galaxies candidates) that come into existence, only a small number of particles will be used to describe the internal structure of many of them, the smaller ones in particular. If this way of proceeding is fully satisfactory from the point of view of a large scale cosmological simulation, it is not viable in the case we want to study in detail a single object of galactic size. This indeed will have total masses ranging from $10^8$ to $10^{13}\, M_\odot$, dimensions from say 1 to several tens of kpc, $M_{DM}/M_{BM}$ in cosmological ratio where DM is by far dominating, and to be properly described will require a large number of particles (at least 50,000) for each component. Repeated zooming in of smaller sub-portions of the whole initial grid is often applied when interested to small size objects such as individual galaxies compared to the cosmological volume.
However, this procedure is quite expensive in terms of computational resources. Other strategies can be found that have already been adopted in some of the studies we referred to. In particular, an isolated perturbation can be artificially created and evolved without following the evolution of the much larger region of space in which it is located. Moreover, since the aim of this study is to explore a wide range of masses and initial densities, the creation of an \textit{ad-hoc} set of haloes is the most appropriate method to  adopt, instead of searching in large scale simulations for the particular haloes suited to our purposes. To this aim, we proceed as follows.

i) We assume the $\Lambda$-CDM concordance cosmology, with values inferred from the WMAP-5 data
\citep{Hinshaw2009}:
flat geometry, $H_0=70.1$ km/s/Mpc, $\Omega_{\Lambda} = 0.721$, $\Omega_b=0.046$ (giving a baryon ratio of $\simeq0.1656$), $\sigma_8=0.817$, and $n=0.96$.

ii) To describe the growth of primordial perturbations, we use COSMICS, the free software written by  \citet{Bertschinger1995}. This is particularly easy to use and it has been already adopted in many studies as the generator of initial conditions for NB simulations. However, we are not interested in a full cosmological simulation containing perturbations at all scales, but only in a portion of it containing a perturbation with properties (over-density, mass, and dimensions) fixed a priori. The size of this sub-portion is fixed in such a way that the wavelength of the perturbation corresponding to the chosen over-density and mass is similar to (but suitably smaller than) the size of the sub-portion. Our aim is to construct a reference proto-halo containing DM and BM particles in the right proportions each of which with its own known mass, position and velocity vector. This reference proto-halo, instead of being constructed by hand as in \citet{Chiosi2002}, is obtained by means of COSMICS in the context of an assigned cosmological scenario. Casting the problem in a different way, instead of searching within a large scale realistic cosmological box the perturbation most suited to our purposes, we suppose that a perturbation with the desired properties is already there, and derive the positions and velocities of all its DM and BM particles from a self-consistent, small-size cosmological box tailored to the perturbation we have chosen.

iv) This reference halo will correspond to the largest mass of the whole sample of galaxy models we have calculated. The grid containing our density perturbation has the size of $l=9.2$ comoving Mpc on a side, and it is populated by a grid of $46^3$ particles. The regular positions of the particles are perturbed by COSMICS  consistently with cosmological random gaussian fluctuations. Furthermore, a density peak is constrained to form at some very early epoch a virialized structure near to the center of the box (COSMICS allows us to specify the properties of the constrained density peak to model the density field in the desired way). We impose a gaussian spherical over-density with average linear density contrast $\delta \rho=3$, smoothed over a region of radius $3.5$ comoving Mpc. We remind the reader that the linear density contrast is defined as $<\rho>/\rho_{bg}-1$ (where $\rho_{bg}$ is the average matter density of the Universe), and it can be extrapolated beyond the linear regime; it would be equal to 1.86 at the epoch of virialization. Requiring a higher value in COSMICS  simply bounces back in time the epoch of virialization.

v) Starting from these input assumptions, COSMICS returns the initial comoving positions and the initial peculiar velocities of the particles at the time in which the highest density perturbation is exiting the linear regime (i.e. it has $\rho/\rho_{bg}-1=1$). This is an important point to remember. COSMICS is used only up to the exit from the linear regime. The subsequent evolution of the DM and BM haloes during the non-linear regime is followed by NB-TSPH simulations.

vi) The small cosmological box containing the perturbation eligible to become a galaxy is our approximation of the local Universe, the one subsequently affecting the evolution of the perturbation itself. The validity of this approximation will be examined in some detail below. To avoid numerical problems with sharp edges, we then single out the sphere of radius $r=l/2$ centered on the center of the box, change the particles coordinates from comoving to the proper physical values (this is simply achieved by dividing the comoving value by the initial cosmological expansion parameter $a=1/z_i - 1$, where $z_i$ is the initial redshift provided by COSMICS at the end of the linear regime), and add a radial outward directed velocity component to each particle. This component of the velocity is  proportional to the radial position of the particle and  the initial redshift of the simulation. This velocity mimics the effects of the outward directed Hubble flow, to take the expansion of the Universe into account.\footnote{Alternatively, the equations of motion could be modified to account for the expansion. However, while this is straightforward to do in a comoving frame, it is less commonly done in physical coordinates. Moreover, virialized structures (like the model galaxies after the initial stages of evolution) are no longer subject to the expansion of the Universe and should be described by static coordinates. Examples of the method we adopt are given in \citet{Katz1991}, \citet{Kawata1999} and in other similar studies.}

vii) To obtain the initial conditions for other model proto-haloes (galaxies) with different mass and/or mean
initial densities, we start from the reference High Density, High Mass (HDHM) halo and proceed as follows:

\noindent \textsf{Varying the mass}. If the spatial dimensions of the box are changed, COSMICS modifies in a self-consistent fashion the size and consequently the mass of the haloes, leaving, however, the mean density unchanged. Playing with this, we generate a number of proto-haloes with the same mean density but smaller mass and consequently radius. Considering  the extremely wide range of galaxy masses observed in the Universe, we generated proto-haloes of different mass whose value differ by a factor of  $\simeq 64$ (and the corresponding radii by factors of 4) passing from one to another. The initial redshifts of models with equal initial density but different mass are different from one another, because the exit time from the linear regime is a function of the mass of the perturbation.

\begin{figure}
\centering
\includegraphics[width=.4\textwidth]{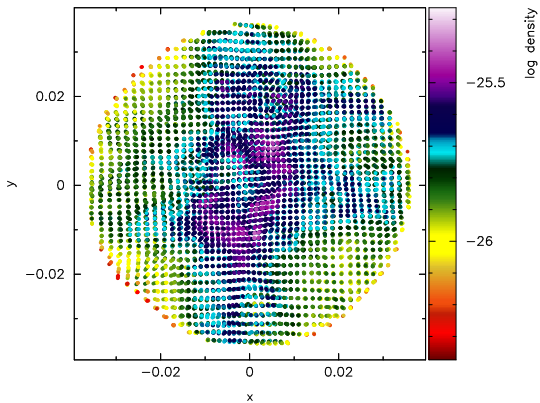}
\caption{Initial displacement of gas particles in a central slice along the $XY$ plane for model IDIM (distances are in proper Mpc). The color code indicates the mass density [g/cm$^3$]. All the other models have similar initial conditions.}\label{ini}
\end{figure}

\noindent \textsf{Varying the density}. The procedure to generate haloes with the same mass but different initial mean density is more complicate. \textit{Before} perturbing the positions (and velocities) of the particles in the grid, we artificially decrease the densities at each grid point of the grid, thus varying the average density but conserving the total mass. Since the initial conditions are produced at the moment in which the highest perturbation peak exits the linear regime, the initial spatial configuration is still a superposition of independent plane waves with different wave-numbers and random phases, i.e. a \textit{gaussian field} with $\delta(\vec{q})=\sum_{\vec{q}}\delta_{\vec{k}}$exp$(i\vec{k}\cdot\vec{q})$, where $\vec{k}$ is the wave-number and $\vec{q}$ is the Lagrangian coordinate corresponding to the unperturbed comoving position of a mass element. Dividing the local (over-)densities $\delta(\vec{q})$ by a constant factor $f$ returns a self-similar gaussian field in which
the Fourier coefficients $\delta_{\vec{k}}$ is divided by the same factor, implying that the variance $\sigma^2 \propto \sum_{\vec{k}} \delta_{\vec{k}}^2$ of the perturbation is reduced by a factor $f^2$. COSMICS subsequently computes the effective displacement and velocities of the particles via the Zel'dovich approximation,

\begin{eqnarray}
&& \vec{x}(\vec{q}) = \vec{q} + D_+ \vec{d}(\vec{q}), \nonumber\\
&& \vec{v}(\vec{q}) = \dot{D_+} \vec{d}(\vec{q}), \nonumber\\
\end{eqnarray}

\noindent where the displacement field $\vec{d}(\vec{q})$ is given by the relation $\vec{\nabla }\cdot\vec{d} =
-D_+^{-1}\frac{\delta\rho(\vec{q})}{\rho}$, and $D_+$ is the cosmic growth factor at the initial redshift, which is a function of the cosmological model. We avoid the simplest choice of imposing a less pronounced density peak (i.e., a lower value of the initial density contrast $\delta$ in the constrained field), because this would reduce the density of the constraint \textit{without} modifying the random perturbations, thus resulting in a very different global field with more spectral power on the small scales. To obtain \textit{intermediate density} haloes, the mean density of the reference halo is decreased by a factor of $15\%$. The mean density of these latter is then decreased  by another $15\%$ to obtain \textit{low density} haloes. Therefore, these latter have the an initial mean density which is $72.25\%$ of the value of the initial reference halo. Finally, a set of haloes with very low initial over-density was obtained reducing the mean density by $50\%$. As already pointed out, the different densities can be considered as a way to mimic the possible environments in which a proto-halo begins to form.

viii) At this stage, a minimal amount of solid-body rotation is also added to all the particles in the proto-haloes. The solid-body rotation is represented by the spin parameter

\begin{equation}
\lambda = \frac{J |E|^{1/2}}{G M^{5/2}}
\end{equation}

\noindent where $J$ is the angular momentum, $E$ is the  initial binding energy,  and $M$ the total mass of the system. Typical values of $\lambda$ range from 0.02 and 0.08 \citep{White1984}, which corresponds to angular velocities of the order of fractions of a complete rotation over time-scales as long as about ten \textit{free-fall} time-scales \citep{Carraro1998}. We adopt $\lambda=0.02$. The choice is motivated by the fact that our model galaxies are meant to represent ETGs. This issue will be also shortly commented in general remarks of this section.

ix) Finally, each particle in each proto-halo is split into a DM and a BM particle (gas), this latter with a small displacement from its original position to avoid numerical divergences. Indicating with $m_0$ the mass of the original particle and with $f_{BD}=\rho_{BM}/\rho_{DM} \simeq 0.1656$ the cosmological ratio between the BM and DM densities, we assign to each gas particle the mass $m_{gas} = f_{BD}\times m_0$. Consequently, $m_{DM} = (1 - 0.1656) \times m_0$. The final HDHM proto-halo consists of $\simeq$ 58,000 DM particles plus an equal number of gas particles.

\noindent \textsf{Coding the models}. With this method, we have obtained twelve initial haloes. As the reference HDHM model, each one is identified by a string of four or five letters, of which the first two (or three) refer to its density and the last two to its mass(e.g., IDLM is for \textit{intermediate density, low mass}; VLDHM stands for \textit{very low density, high mass}, etc.). The key initial value of a few important parameters for all the twelve haloes are listed in Table \ref{tab1}, whereas a sketch of the density contrast inside the initial sphere projected on the $XY$ plane for the case of the model IDIM is shown in Fig. \ref{ini} where a number of lumps of matter are clearly evident. The number of particles in all models is similar to that of the HDHM case, with some minor differences due to the different displacement of the particles in the outskirts of the retailed sphere.

\textsf{General remarks}. We point out that all the proto-haloes we have chosen are fully consistent with the underlying cosmological background. The expected DM halo mass function follows the laws by, e.g., \citet{Press1974} or \citet{Sheth2001}, and/or the analytical fits of numerical simulations \citep[e.g.][]{Warren2006}. In this picture, a $M \simeq 10^{13} M_{\odot}$ halo is becoming ``typical'' on a $\sim 10$ Mpc scale (the typical linear size of a large galaxy cluster) at redshifts below $z\sim 5$, which is compatible with the collapsing redshift of our  HDHM  model (see below). Such massive haloes are also expected not to be rare on ten times larger scales at $z\sim 10$. This can be easily checked considering the halo growth functions of DM haloes for the concordance $\Lambda$-CDM cosmology: the expected number of haloes with mass $\simeq 10^{13} M_{\odot}$ within a volume of 1 (Mpc$/h)^3$ at $z \sim 10$ is $n \simeq 10^{-8}$  \citep[see Fig.2 of][]{Lukic2007}.

The main criticism that could be made to the above procedure is that the size  of the simulation backing the perturbation in question is too small (9.3 comoving Mpc on a side for the HDHM box, and consequently smaller in the other cases). It is long known that the simulation size determines the maximum perturbation wavelength. If the long wavelengths are dropped out, the strength of subsequent clustering is reduced, but at the same time the number density of intermediate mass haloes (with total mass of the order of $10^{13} h^{-1} M_\odot$ in our case) is enhanced \citep{PowerKnebe2006}. While this may be a problem in real cosmological simulations, in our case it is not so, because we are interested in objects with the mass and size of galaxies and not of galaxy clusters. Furthermore, \citet{PowerKnebe2006} show that truncation of the initial power spectrum (i.e., the small size) of the simulation has little impact on the internal properties of the haloes. However, truncation lead to spin parameters that are 15\% lower than usual. We have in a sense avoided the whole spin problem just by taking the lowest value for the spin parameter ($\lambda=0.02$) suggested by \citep{White1984}, a reasonable approximation for slowly rotating systems such as ETGs. A final remark could be made related to the fact that adopting the sphere of radius $r=l/2$ (about the size of the Local Group) as the starting proto-halo we implicitly neglect the possibility that late infall of nearby  haloes that initially were outside the sphere may occur. This would inhibit late refuelling of gas (and stars and DM) to our system that is equivalent to say that we inhibit late mergers likely accompanied by  star formation. This indeed is not the aim of the present study,  which  intends to explore the modality of star formation in alternative to the very popular hierarchical scheme. To conclude, the scheme we propose to derive the initial conditions for our model galaxies is not in conflict with the cosmological paradigm and it is easy to implement in NB-TSPH simulations.

\begin{table*}
\caption{Initial parameters for the twelve models of proto-galaxies. Left to right: total masses, corresponding (initial) gaseous masses, mean halo over-densities ($\delta \rho \def \rho / \rho_{background} -1$) at redshift $z=30$, initial  redshift, halo initial  proper physical radius, mass of a gas particle.} \centering
\begin{tabular}{|l|l|l|l|l|l|l|}
\hline
Model & $M_{tot}$ [$M_{\odot}$] & $M_{gas,ini}$ [$M_{\odot}$] & $<\delta\rho-1>_{z=30}$ & $z_{ini}$ & $r_{ini}$ [kpc] & $m_{gas}$ [$M_{\odot}$] \\
\hline
\hline
HDHM & $1.75\times10^{13}$ & $2.90\times10^{12}$ & 0.39 & 46.34 & 97.17  & $4.97\times10^{7}$ \\
\hline
IDHM & $1.75\times10^{13}$ & $2.90\times10^{12}$ & 0.30 & 39.24 & 114.31 & $4.97\times10^{7}$ \\
\hline
LDHM & $1.75\times10^{13}$ & $2.90\times10^{12}$ & 0.23 & 33.20 & 134.49 & $4.97\times10^{7}$ \\
\hline
VLDHM & $1.75\times10^{13}$ & $2.90\times10^{12}$ & / & 22.67 & 194.34 & $4.97\times10^{7}$ \\
\hline
HDMM & $2.69\times10^{11}$ & $4.45\times10^{10}$ & 0.46 & 53.79 & 20.99 & $7.79\times10^{5}$ \\
\hline
IDIM & $2.69\times10^{11}$ & $4.45\times10^{10}$ & 0.33 & 45.57 & 24.69 & $7.79\times10^{5}$ \\
\hline
LDIM & $2.69\times10^{11}$ & $4.45\times10^{10}$ & 0.25 & 38.59 & 29.05 & $7.79\times10^{5}$ \\
\hline
VLDIM & $2.69\times10^{11}$ & $4.45\times10^{10}$ & / & 26.37 & 41.98 & $7.79\times10^{5}$ \\
\hline
HDLM & $4.17\times10^{9}$ & $6.91\times10^8$ & 0.54 & 63.23 & 4.48 & $1.22\times10^{4}$ \\
\hline
IDLM & $4.17\times10^{9}$ & $6.91\times10^8$  & 0.39 & 53.60 & 5.27 & $1.22\times10^{4}$ \\
\hline
LDLM & $4.17\times10^{9}$ & $6.91\times10^8$  & 0.29 & 45.40 & 6.20 & $1.22\times10^{4}$ \\
\hline
VLDLM & $4.17\times10^{9}$ & $6.91\times10^8$  & 0.16 & 31.11 & 8.96 & $1.22\times10^{4}$ \\
\hline
\end{tabular}
\label{tab1}
\end{table*}

\subsection{The NB-TSPH code \textsc{EvoL}}

\textsc{EvoL} is a Lagrangian N-Body parallel code designed to study the evolution of astrophysical systems on any spatial scale, from large cosmological volumes to sub-galactic regions. It is based on the standard Tree Algorithm for the description of the gravitational interactions \citet{Barnes1986}, and on a modern version of the Smoothed Particle Hydrodynamics method to model gas dynamics. Its basic features concerning the treatment of gravitational interaction and hydrodynamics are described in details in \citet{Merlin2009} and \citet{Merlin2010}; here,
we briefly recall some of its characteristics.

\textsc{EvoL} includes some peculiar features, such as the lagrangian formulation of the SPH method that includes the so-called $\nabla h$ terms in the equations of motion \citep[taking into account variation of the
smoothing length, see][ for the definition]{Merlin2010}, an artificial thermal conduction to smooth out
contact surface discontinuities, and finally the variable adaptive softening lengths with consistent correcting terms in the equations of motion. In the present models we make the following assumptions and/or simplifications:

(i) We switch off the artificial thermal conduction, because its effects in combination with radiative cooling are not yet fully understood and tested, and may lead to unphysical behaviors, with particles exchanging heat because of
numerical conduction and simultaneously losing energy due to physical cooling. The issue deserves an accurate analysis to be planned  for a forthcoming study.

(ii) We adopt variable softening lengths; this is an important difference with respect to other similar calculations, in which  a constant softening length has been adopted. The only exception is  \citet[][]{Nelson2006}
even if  in that case no correcting terms in the equations of motion were considered. In the present models, we let the softening, $\epsilon$, and smoothing, $h$,  lengths vary freely. In a few companion  models calculated for the sake of
comparison, we adopt for the two parameters the minimum threshold values $\epsilon_{min}=100$ and $h_{min}=100$ pc. Such low limits would provide sufficient resolution in all the cases of interest. However, these side models show that the results are very sensitive to these parameters, even if all the models are essentially similar in their
global aspects (see Sect. \ref{reliable} for a discussion on this issue).

(iii) We include a density-dependent pressure limit to avoid artificial clumping of poorly resolved gas. The method is similar to the one presented by \citet{Robertson2008}. In practice, at each time-step all SPH particles are used to compute the local Jeans mass, which is a function of density and temperature, i.e.

\begin{eqnarray}
m_{Jeans} = \frac{4 \pi}{3} \rho_{gas} \left( c_s \sqrt{\frac{\pi}{G
\rho_{tot}}} \right)^3,
\end{eqnarray}

\noindent where $c_s = \sqrt{\gamma (\gamma-1) u}$ is the local sound speed, $\rho_{gas}$ is the density of the SPH particle, and $\rho_{tot}$ is the \textit{total} local density (i.e., gas plus DM and/or stars)\footnote{This expression for the Jeans mass of a gas sphere in presence of a non collisional component can be obtained from a perturbative analysis of the stability of a pressurized cloud, see e.g. \citet{Tittley2000} for reference.}.
Then, the internal energy of each particle (limited to the computation of the hydrodynamical forces) is chosen as the maximum between the real value and the ``effective'' value, which is defined as the energy that the gas sphere should have to be stabilized against collapse:

\begin{eqnarray}
u_{eff} = N_{jeans}\times\frac{1}{\gamma(\gamma-1)}\rho_{tot}
\left( \frac{3N_{SPH}m}{4 \pi^{\frac{5}{2}} \rho_{gas}} \right)^{\frac{2}{3}},
\end{eqnarray}

\noindent where $m$ is the particle mass, $N_{SPH}$ is the number of particles required to resolve a region (indicatively, this number can be considered equal to the typical number of neighbours, i.e. about 60 in our  models, or  a multiple of this value), $\gamma$ is the adiabatic index, and $N_{jeans}$ is a free parameter $\geq 1$.
In these models we use $N_{jeans}=15$ \citep[][in preparation]{Merlin2009,Merlin2012}. Note that $u_{eff}$ is only used to compute the mechanical acceleration due to the internal gas pressure. In the expressions for the variation of internal energy, dissipation, and cooling, the \textit{real} value of $u$ is  adopted. In this way, the temperature of gas particles can reach low values, without reaching high densities, thus creating regions of cold material
without spurious numerical clumping of particles.

(iv) We run the simulations in physical coordinates since at present we are not interested in large scale motions. Moreover, this choice rules out the need for periodic boundary conditions, which would require a large amount of additional CPU-time.

(v) Our initial system is confined within the initial sphere of $l/2$ radius and it is surrounded by empty space.
In other words the evolution of our model galaxy occurs  \textit{in void}. This simplification of the problem  has
some non-negligible drawbacks. First of all, the infall of material towards to innermost regions of the galaxy sooner or later stops because material outside the sphere cannot flow in. This rules out the possibility of late mergers and/or late infall of significant amounts of gas. Important point to keep in mind, but which does not invalidate the results of this study as long as late merger, matter acquisition are not of interest.  Secondly, the
feedback-heated gas leaving the galaxy will escape freely without experiencing drag forces. Thirdly, the density of the system cannot be consistently computed and the models evolve as small closed Universes rather than over-dense regions within a flat Universe, so the redshifts are only indicative of the real evolutionary time-line.

(vi) Finally, we include radiative cooling, star formation, and stellar feedback. As they are described in detail in
\citet{Merlin2009}, in the sections below we limit ourselves to summarize their key aspects.

\subsection{Radiative cooling}\label{cool}

The cooling functions for atomic radiative processes are those described in \citet{Carraro1998}, which in turn are based on those elaborated by \citet{Chiosi1998}. In brief, for temperatures greater than $10^4$ K they lean on the
\citet{Sutherland1993} tabulations for a plasma under equilibrium conditions and metal abundances $\log [Z/Z_{\odot}]$ = -10 (no metals), -3, -2, -1.5, -1, -0.5, 0 (solar) and 0.5. The chemical properties of gas particles are followed in details, tracking the abundances of Fe, O, Mg, Si and C, and the global metallicity Z. For temperatures in the range $100<$T$<10^4$ the dominant source of cooling is the $H_2$ molecule becoming rotationally and/or vibrationally excited through a collision with an H atom or another $H_2$ molecule and decaying through radiative emission; the data in
use have been derived from the analytical expressions of \citet{Hollenbach1979} and \citet{Tegmark1997}, amalgamated together by \citet{Chiosi1998}. We point out that in the present simulations it is assumed that \textit{all} hydrogen
becomes molecular below $10^4$ K, so the cooling rate in this range of temperatures may be somewhat overestimated. However, cooling from collisions of neutral hydrogen atoms with electrons from ionized metals (when present) is neglected. This may somewhat compensate  for the excess of cooling due to molecular hydrogen. In any case this is a (marginally) weak point of our  current treatment of cooling that should be improved.

Finally, for temperatures lower than 100 K, \citet{Chiosi1998} and \citet{Carraro1998}, starting from the studies of \citet{CaimmiSecco1986} and \citet{Theis1992}, incorporate the results of \citet{Hollenbach1979} and \citet{Hollenbach1988} for CO molecule as the dominant coolant. The following analytical relation in which
the mean fractionary abundance of CO is given as a function of [Fe/H], is found to fairly represent the normalized cooling rate (i.e. $\Lambda_{CO}/ n^2$ with $n$ the number density of particles)

\begin{equation}
 {\Lambda_{CO} \over n^2} = 1.6 \times 10^{-29}10^{([Fe/H]-1.699)T^{0.5}} \, \,
\end{equation}
\noindent in  erg/s/cm$^{-3}$.  For all other details see \citet[][]{Carraro1998}.

We also include the contribution by inverse Compton cooling
\begin{eqnarray}
&&\Lambda_{Compton} = 5.41\times10^{-36} x_e [T - T_{CMB}(1+z)] (1+z)^4 \nonumber\\
&&
\end{eqnarray}

\noindent in erg/s/cm$^3$, with the usual meaning of symbols
\citep[see e.g.][]{Ikeuchi1986}.

The radiative cooling is computed separately from the SPH equation of energy conservation, due to its very short time-scales. During a dynamical time-step, the density and the metallicity of gas particles are kept fixed, as well as the mechanical heating rate. On the contrary, the temperature is let vary under the action of cooling, and the new
cooling rates are simultaneously obtained as a function of the new temperature. The process is iterated until the end of the time-step.

\subsection{Star formation} \label{starfo}

Star formation is modeled by means of a stochastic method, similar to that introduced by \citet{Churches2001} and \citet{Lia2002}. First, only gas particles belonging to convergent flows (i.e. $\nabla \cdot v <0$) and denser than a suitable threshold $\rho_*$ are considered eligible to form stars. In our models we adopt $\rho_* = 5\times10^{-25}$ g/cm$^3$. No restriction on the temperature is imposed; this is motivated by the fact that thermal instabilities can produce star forming sites even within high temperature gas.

If a gas particle satisfies these criteria, it is assumed to form stars at the rate $d\rho_*/dt =  \epsilon_{SF}\rho_g/t_{ff}$, where $t_{ff}\simeq 0.5/\sqrt{G \rho_{tot}}$ is the free-fall time, $\rho_{tot}$ is the local total mass density (DM plus BM), and $\epsilon_{SF}$ the dimensionless efficiency of the star formation
process. This means that a gas particle is expected to transform a fraction $\epsilon_{SF}$ of its mass into stars over its free-fall time scale. However, a stochastic description of the star forming process is adopted to avoid the creation of exceedingly large numbers of star particles. Thus, gas particles undergo a Monte Carlo selection to check whether or not they will actually form stars, in such a case they are instantaneously turned into collision-less star particles \citep[see also][]{Lia2002, Merlin2009}. To this aim, a random number $r \in [0,1]$ is drawn and compared to the probability that the particle is actually forming stars. An obvious choice would be to use the probability $P = \epsilon_{SF} / t_{ff} \times \Delta t$, where $\Delta t$ is the dynamical time-step. However, the evaluation of the real probability must take into account that the dynamical time-steps are generally much shorter than the free-fall times of particles. This implies that within a free-fall time scale a number of star forming events and corresponding random draws are possible depending  on $\Delta t$, and therefore so does the global probability for a single gas particle to be turned into stars. If the global probability over $t_{ff}$ is $P$, the stochastic process of random selection must be corrected by considering that the probability summed over the number of draws $n \simeq \Delta t / t_{ff}$ must be equal to the probability of a single draw within the whole free-fall time. First we take
as the probability within a single time-step the value $p \simeq P/n$, then we  consider as the probability of a successful event over $n$ draws the quantity $p \times \sum_{i=1}^n (1-p)^{i-1} < P$. The correcting multiplicative term is

\begin{equation}
f=\frac{P}{\frac{P}{n} \times \sum_{i=1}^n (1-\frac{P}{n})^{i-1}}.
\end{equation}

\noindent Therefore, the final probability during the time-step is $f \times p$. If $r \leq fp$, then the gas particle is instantaneously turned into a stellar particle.

In the present models, we assume $\epsilon_{SF} = 1$. Empirical estimates of the efficiency of star formation based on observational data of star forming events inside molecular clouds in the local vicinity yield  $\epsilon_{SF} \simeq 0.025$ \citep{Lada2003, Krumholz2007}. In this case the gas density and free-fall time scale of are those of the cold molecular clouds. However, in the case of the large scale star formation mechanism in a galaxy, the above estimate may not correspond to reality. As a matter of facts, considering that a typical galaxy with $10^{11} M_\odot$ mass in stars has to disposal a time scale of about 13 Gyr to build up its stellar content, the mean estimate of the star formation efficiency is closer to 0.1 rather than 0.02. The new estimate would increase by nearly a factor of ten if the time to disposal to form stars is much shorter than 13 Gyr, say 1 to 2 Gyr. In addition to this, it is not known whether $\epsilon_{SF}$ should be maintained constant during different cosmic epochs due to the different mechanisms forming stars at high redshifts and in metal poor regions with respect to local molecular clouds. Finally, it is not known whether the efficiency of star formation is the same in all galaxies independently of their mass (either total or baryonic). For all these reasons, we prefer to adopt here $\epsilon_{SF}=1$, which means that the process of star formation occurs at 100\% efficiency (meaning that \textit{all} gaseous particles satisfying the SF criteria will turn their total  mass into stars within a free-fall time scale). Another consideration has to be made here. Several numerical simulations calculated with different values of $\epsilon_{SF}$  clarify that within a certain range of values, the star formation histories of the models with small values of $\epsilon_{SF}$ are apparently quite similar to those with $\epsilon_{SF} = 1$, the only difference being that with small values of $\epsilon_{SF}$ the simulations become much more time-consuming. The reason is that dense and cold clumps of matter, interacting with hotter material heated up by close SN explosions, require extremely small time-steps. In contrast, if the gas can easily form stars, this critical situation is avoided, and the simulations proceed much faster. The weak impact of $\epsilon_{SF}$ on the final SF history can be attributed to the self-regulation cycle between star formation and energy feedback. A high SF efficiency implies a strong and sudden burst of stellar activity; but young stars soon pressurize their surroundings via energy feedback (see below), quenching the formation of new stars. However, if the gas is sufficiently dense, radiative cooling is effective and further star formation can soon take place (positive feedback). With a low efficiency, less stars form. Their heating is consequently lower, and more stars can soon form \textit{before} the feedback halts their formation. This ultimately leads to the same situation as in the previous case, with perhaps the minor consequence of a delayed enrichment in heavy elements of the medium. Clearly, other parameters play a more important role; for example, the density threshold $\rho_*$ and the efficiency of feedback. Therefore the choice $\epsilon_{SF}=1$ and fast calculations likely, is the best compromise. However, one should always keep in mind that adopting $\epsilon_{SF}=1$ may have other consequences on the dynamical evolution of the systems, favouring the collision-less collapse instead of the dissipative one (see Sect. \ref{conclusions}).

\subsection{Energy feedback}\label{feedback}

When a gas particle is turned into a star particle, it can be considered to represent a Single Stellar Population (SSP) made of many real stars. It starts re-fueling the ISM with heavy chemical elements and energy, mainly because of winds from young massive stars and Supernova (SN) explosions.

\textbf{Winds}. We consider two regimes for the energy injection by winds. When the SSP is young, the main source of energy are the young massive stars. Their winds have very high speeds, typically from 1000 to 3000 km/s. Here we adopt
a constant velocity of $v_{YSO}=1500$ km/s.  The kinetic energy of the winds is assumed to be thermalized and released within the surrounding medium with an efficiency of 20\% \citep[see e.g.][]{Dyson1997}. In the late stages of the SSP evolution, slow velocity winds from low mass stars become predominant; in these case we assume the wind speed of
$v_{OSO}=60$ km/s (typical of RGB stars), and the same thermalization efficiency.

\textbf{SN explosions}. Each \textit{real} SN explosion is expected to deposit some $10^{51}$ kinetic ergs in a very small region, and on a short time scale. However, most of this energy is soon radiated away, and only a small fraction of it is subsequently thermalized. We use the results by \citet{Thornton1998} to obtain an analytic approximation of the fraction of the initial energy which becomes available at the end of the expanding phase of the SN bubble. However, in the real Universe SN explosions take place in already shock-heated regions, with temperatures raised up to some $10^6$ K, because of the action of a photo-ionizing flux from massive young stars, stellar winds, and previous SN explosions. All this is not taken into consideration in Thornton's study, so a large  uncertainty still affects the
description of the whole process. However there are a few exceptions like \citet[][]{Cho2008} that should be taken into account. A single star particle represents an entire SSP, so a large number of SN explosions are expected to take place within a single star particle over a rather long time scale. The number of SN explosions (attention must be paid to release the energy in discrete bursts consistently with the number of real explosions), as well as the amount of gas released by a SSP of given age and mass during a time-step and its chemical composition are computed with the technique described in \citet{Lia2002}, using theoretical SSPs \citep[in \textsc{EvoL}, the Padova tracks are adopted, see][]{Greggio1983}, and adopting the initial mass function for the real stars in the SSP (star particle)
of \citet{Kroupa2001}. We refer the reader to the above papers for all the details. It is worth clarifying that we do \textit{not} adopt stochastic approach of \citet{Lia2002} to model SN explosions. At each time-step  they decided the fate  of each star particle by means of the Monte-Carlo method; randomly chosen  star particles were instantaneously turned back into gas, while the remaining  ones continued their life as  star particles. On the contrary, in our
description each SSP continuously releases gas and energy at each time step, becoming a \textit{hybrid} particle (in the sense described below).

\begin{figure} 
\centering
\includegraphics[width=.4\textwidth]{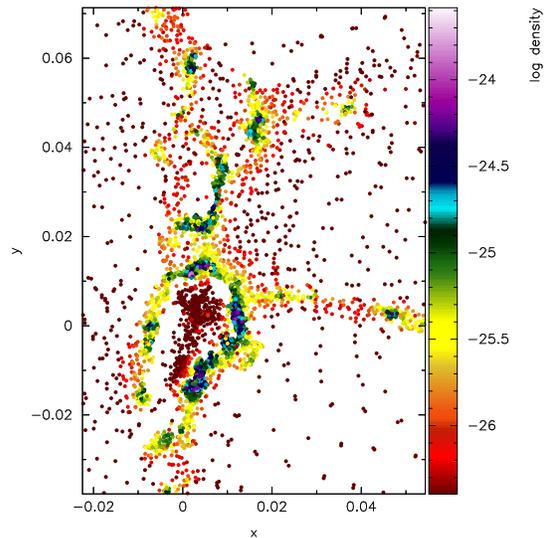}
\caption{Snapshot showing the displacement of gas particles on a central slice on the $XY$ plane (model IDIM, $z=7.6$; distances are in proper Mpc). Features caused by energy injections from stars (the clumped brown particles near the origin of coordinates) such as expanding bubbles and shock fronts  as well as collapsing cosmological filaments are clearly visible.} \label{gasbubble}
\end{figure}

Finally, the total budget of thermal energy produced by stellar winds and SN explosions is given continuously to a sort of gas reservoir assigned to star particles, ultimately formed by the gaseous material ejected from the SSP. This reservoir, due to its high temperature, acts a piston on the neighboring particles, effectively injecting kinetic energy into the surrounding interstellar medium. In the next Section we clarify some aspects of  this method.

\subsection{Fate of the gas ejected by SSPs}\label{gas_eje}

The gas ejected by SSPs, heated up by  the energy budget from winds and SN explosions and enriched in chemical elements, interacts with the surrounding medium. This is a crucial issue in numerical simulations of galaxy formation.
Since it is not feasible to continuously form new gas particles from the stellar particles, it is necessary to develop a consistent, yet simple method to model this process.

There are two main sources of uncertainty. First, one has to decide how to deal with the hot gas \textit{soon after} its ejection from the SSPs (this gas is essentially composed by matter ejected by stars in form of winds and/or SN explosions). How should it be let cool? How to compute its density? It has long been known that distributing the feedback energy among neighboring particles and then letting them normally evolve among the other cold ones yields
poor results and very low feedback efficiencies \citep{Katz1992}: the energy is soon radiated away and, in critical situations, hot particles escape from the region due to excessive hydrodynamical friction with their neighbors that is not physically grounded. Also, distributing the gas released by the SSPs to nearby SPH particles at each time-step has proven to be inefficient, since the large amounts of energy released locally are smoothed out too strongly. To cope with all this, we model we proceed as follows. We suppose that initially  the ejected gas remains stuck to the parent star particles. This assumption sounds reasonable in very early stages of gas ejection, but surely becomes unphysical as time goes by. Sooner or later, the gas ejected by SSPs should move freely and mix with its nearby particles. Therefore, the issue is  how and when the gas should be let evolve ``normally'', as a standard SPH gas particle - i.e., when it should dynamically decouple from its parent star particle.

To this aim, we split the evolution of the ejected gas into two temporal steps. The gas is kept locked to the parent SSP during the first stages of its evolution, namely from its first appearance in the SSP to the age of $t_{eject}\simeq8\times10^7$ years. This is the typical lifetime of a $6\,M_\odot$ star with solar composition and convective overshooting \citep{Bertelli_etal_1994}. In this type of stellar models, the $6\,M_\odot$ star is also the lower limit for the occurrence of Type II Supernovae. The value of $t_{eject}$ could be lowered by a factor of 2 (and the mass limit accordingly increased to about $8-9 \, M_\odot$) to be more conservative, however without changing the overall picture we are proposing. During this first stage, the ejected gas does not contribute to the acceleration of the \textit{parent} particle, but acts as a piston exerting pressure on the neighbouring particles because of its high internal energy. Its density is computed normally with the SPH method, and radiative cooling is let occur normally during this phase. In this way, this ``almost-zero-mass'' SPH particle, locked to a collision-less star particle, can effectively act as a source of feedback, pressurizing the region and giving rise to complex structures resembling observed bubbles and super-bubbles as shown in Fig. \ref{gasbubble}. The mechanism is similar to the one proposed by
\citet{Pelupessy2005}, who used zero-mass \textit{pressure particles} to model the stellar feedback. However, instead of introducing zero-mass particles, in \textsc{EvoL} the real gaseous mass ejected from SSPs during the current time-step (always much smaller than the typical mass of a SPH particle) is computed and added to a gaseous reservoir locked to the star  particles; these ``hybrid'' particles interact hydrodynamically with the surrounding gas particles via the pressure exerted by their gaseous component.

When $t_{eject}$ is elapsed, the hybrid star/gas particle is let free  to collect a sufficient amount of gas (ejected by nearby SSPs in their first stage  too)  from other neighboring hybrid particles  to form a whole new gas particle. When the mass of the collected gas  is equal or comparable to the mass of a standard SPH particle  (in the present models we allow  $\sim 50\%$ tolerance), then a new gaseous particle is spawned, with position, velocity and temperature obtained from the  weighed mean over all parent star particles that contributed to create it with their gas supply (the mass of gas in these hybrid particles is accordingly decreased by the amount of gas they have given away). From now on, the new particle is let free to evolve normally. This method has proven to give satisfactory results, at the expense of a slight increase of the total number of particles during the run.

\begin{figure}
\centering
\includegraphics[width=.4\textwidth]{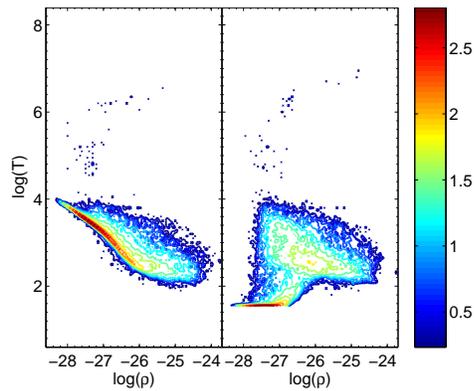}
\caption{Phase diagram $\rho$ (g/cm$^3$) vs. T (K) for the IDLM model at $z=8.5$, with (left) and without (right) the inclusion of a photo-ionizing background radiation at $z=9$. The lines are the iso-levels of the logarithm of the number of particles at each point of the phase diagram, as indicated by the color scale. See text for details.}\label{sfrreion}
\end{figure}

\subsection{Chemical enrichment}\label{chem_enri}

Stellar gaseous ejecta contain heavy elements, which are redistributed within the surrounding gas by means of  a diffusive process, similar to the diffusive approximation adopted for the thermal conduction. In practice, the amounts of heavy elements (namely, Fe, Mg, O, Si, C, and global metallicity Z) ejected by each SSP, i.e. star particle, is computed at each time-step by means of the Padova evolutionary tracks and the \citet{Greggio1983} rates based on  the number of SN events and the amount of metals released by the ``dying'' SSPs during the current time-step. The global chemical yield from a single SSP during each time-step is given by the elements released by dying (exploding) stars during that time-step. These are the sum of three contributions: (i) the elements locked in the dying stars at the moment of their birth, inherited from the proto-stellar gas; (ii) the heavy elements created in the core  by thermonuclear reactions and ejected in Supernova II explosions; (iii) the elements created in the core of binary stars which end their life as Supernovae Ia. The reader should refer to \citet{Lia2002} for a detailed description of the method.

Numerically, the heavy elements released by a SSP are  assigned to the gaseous mass ejected by the same SSP during the current time-step and subsequently spread over neighboring particles by means of a diffusion algorithm based on the Fick's  law in spherical symmetry:

\begin{equation}
\frac{\partial f}{\partial t} = D_{diff} \frac{\partial^2 f}{\partial r^2}.
\end{equation}

The formalisms is the same we have adopted for thermal conduction due to diffusion of electrons. The coefficient $D_{diff}$  is obtained from the diffusion velocity of electrons by \citet{Monaghan1992}.

\begin{figure}
\centering
\includegraphics[width=.4\textwidth]{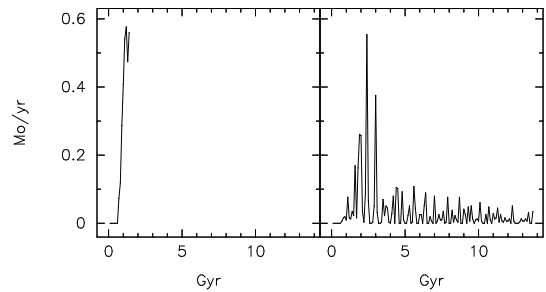}
\caption{Star formation histories of two low mass models with (left) and without (right) energy feedback from the
stars.}\label{sfrNoF}
\end{figure}

\subsection{Re-ionization and the first generation of stars} \label{reion}

Cosmic re-ionization, as inferred from the WMAP-5 data, is currently believed to have taken place at $z \leq 10$. In our models, we include a crude treatment of photo-heating from  re-ionized free electrons. Following \citet{Pawlik2009}, we increase the temperature of all gas particles fueling their internal energy reservoir with an
additional budget of 2 eV per proton, which is instantaneously injected at $z=9$. The effect of this energy boost is to raise the temperature of the low density particles to some $\sim 10^4$ K. In high density regions, however, radiative cooling is so  efficient to keep the temperature well below this value. Fig. \ref{sfrreion} compares two snapshots taken from models in which this re-ionization effect is (is not) included. However, it is worth noting that this way of proceeding lets the gas particles free to cool down soon after the sudden injection of energy; consequently, the cosmic web may happen to get too cold at later times (this can be seen in Fig. \ref{rhoTz2} of Sect. \ref{accrete}).

In this context, we also include in a very simple fashion an early population of metal-free stars otherwise known as Population III (Pop III) stars \citep[e.g.,][\, and references therein for more details]{Bromm2002,Yoshida2003}. We assume that when a metal-free gas particle is prone to form stars for the first time (see Sec. \ref{starfo}) it generates only Pop III objects. Furthermore, this  SSP is supposed to contain only massive stars, because according to current believe the underlying initial mass function is  heavily skewed towards massive stars \citep{Bromm2002}. Consequently the Pop III generation of stars is short lived and within a few million years (almost immediately on galactic time scales) injects lots of energy into the interstellar medium by stellar winds and supernova explosions. Therefore, thanks to this we can simplify the occurrence of this first generation as follows: gas particles undergoing star formation are not turned into stars, but simply raised to high temperature, $T\simeq 2\times10^7$ K, and enriched in metals, $Z\simeq 10^{-4}$. By doing this we do not care to track the fate of Pop III stars as far as their remnants (ultra-compact objects and Black Holes) are concerned.  Since these remnants could be the seeds to form to a central super-massive black hole, most likely fueling a period of AGN activity, a complete description of the remnants from Pop III stars is in progress.

\begin{figure*}
\centering
\includegraphics[width=14cm,height=3.5cm]{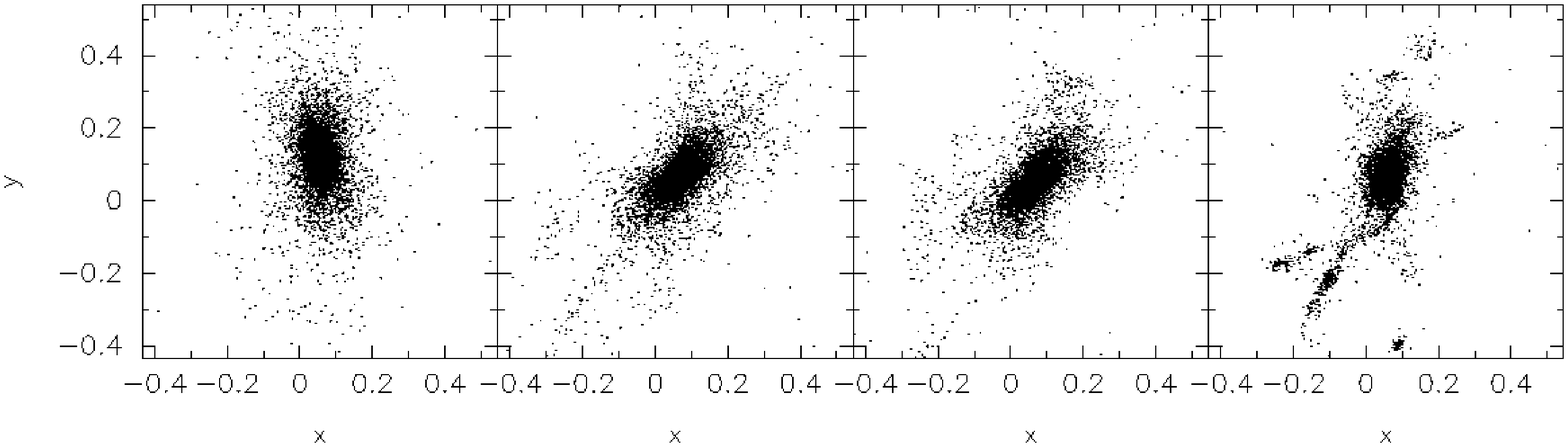}
\includegraphics[width=14cm,height=3.5cm]{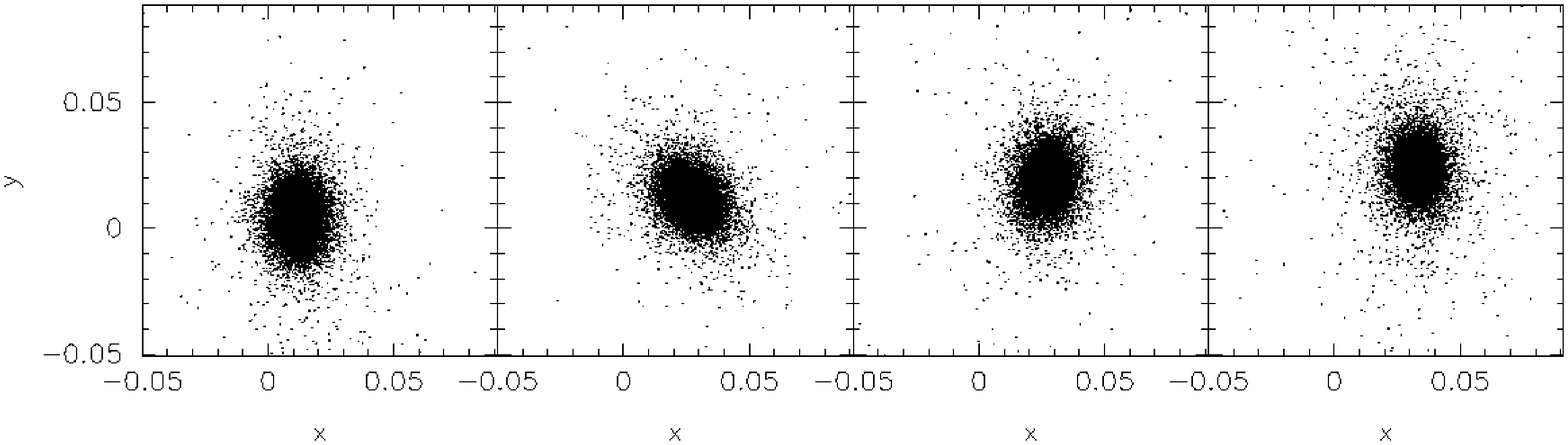}
\includegraphics[width=14cm,height=3.5cm]{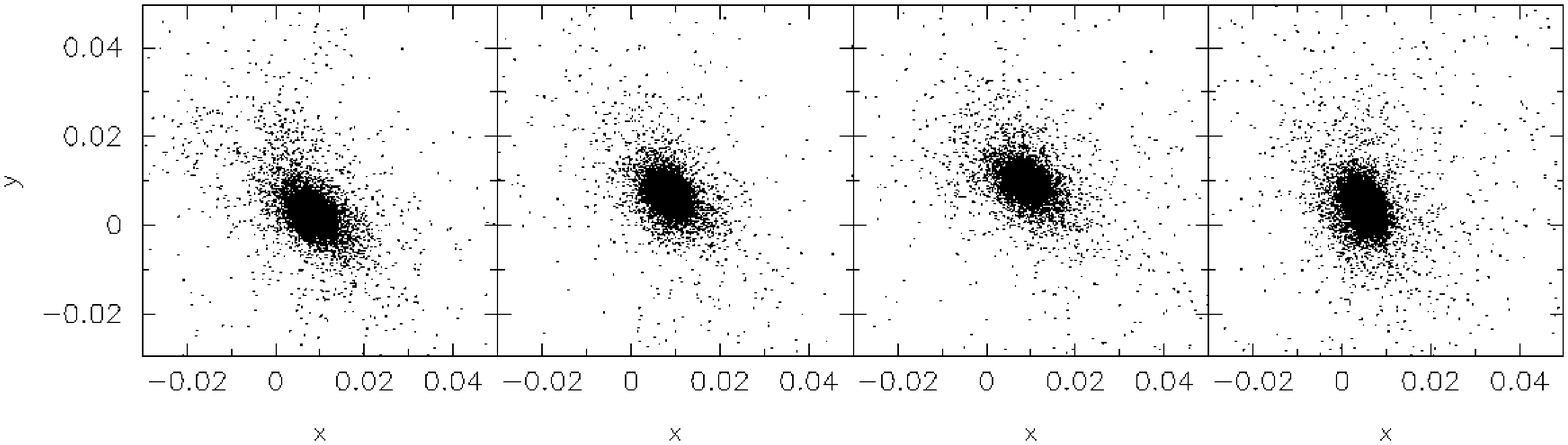}
\caption{The spatial distribution of the stellar content of the model galaxies at the redshift of the last computed stage, projected on the
$XY$ plane. Left to right, top to bottom: HDHM ($z=0.22$), IDHM ($z=0.77$), LDHM ($z=0.49$), VLDHM ($z=0.83$);  HDIM ($z=1.0$), IDIM ($z=0.75$), LDIM ($z=0.58$), VLDIM ($z=0.15$); HDLM ($z=0.36$), IDLM ($z=0.22$), LDLM ($z=0.05$), VLDLM ($z=0.0$). The coordinates X and Y are  in proper Mpc.}\label{starsb}
\end{figure*}

\begin{figure*}
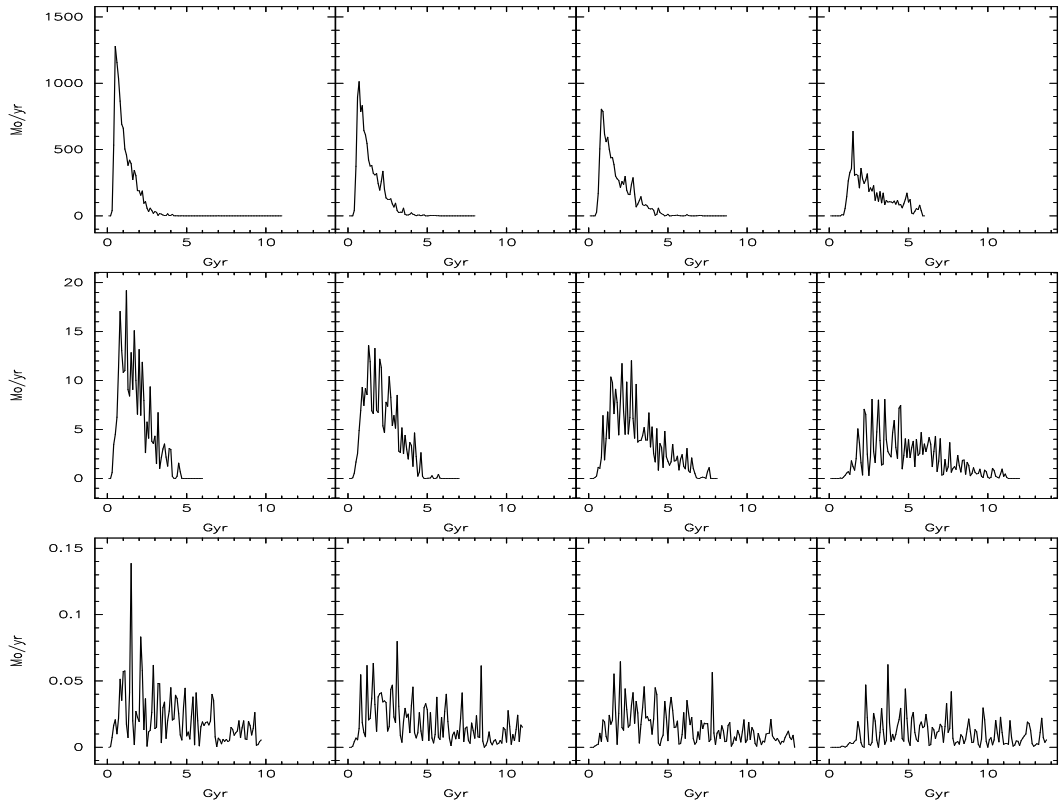

\centering
\includegraphics[width=14cm,height=3.5cm]{Figures/sfr1.ps}
\includegraphics[width=14cm,height=3.5cm]{Figures/sfr2.ps}
\includegraphics[width=14cm,height=3.5cm]{Figures/sfr3giuste.ps}
\caption{Star formation histories for the twelve reference models. Left to right: high density, intermediate density, low density, very low density. Top row: high mass; intermediate row: intermediate mass; bottom row: low mass.}\label{sfr2}
\end{figure*}

\begin{figure*}
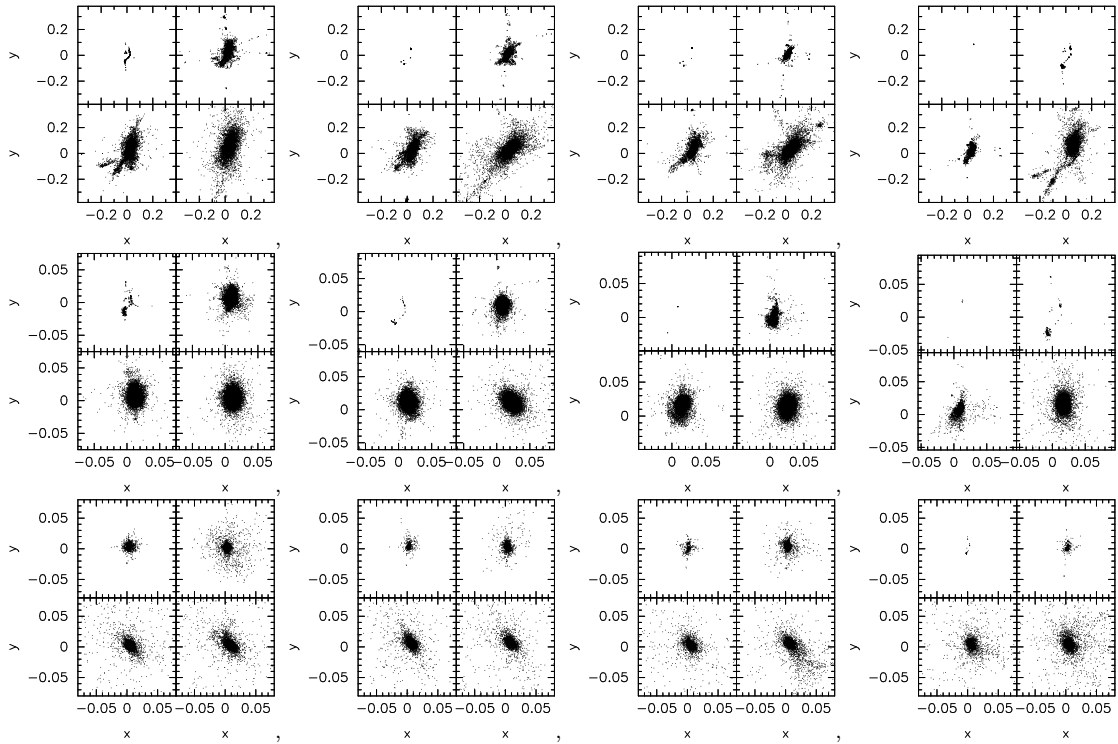

\centering
{\includegraphics[width=.2\textwidth]{Figures/HDHMstarEv.ps},
\includegraphics[width=.2\textwidth]{Figures/MDHMstarEv.ps},
\includegraphics[width=.2\textwidth]{Figures/LDHMstarEv.ps},
\includegraphics[width=.2\textwidth]{Figures/VLDHMstarEv.ps} }

{\includegraphics[width=.2\textwidth]{Figures/HDMMstarEv.ps},
\includegraphics[width=.2\textwidth]{Figures/MDMMstarEv.ps},
\includegraphics[width=.2\textwidth]{Figures/LDMMstarEv.ps},
\includegraphics[width=.2\textwidth]{Figures/VLDMMstarEv.ps} }

{\includegraphics[width=.2\textwidth]{Figures/HDLMstarEv.ps},
\includegraphics[width=.2\textwidth]{Figures/MDLMstarEv.ps},
\includegraphics[width=.2\textwidth]{Figures/LDLMstarEv.ps},
\includegraphics[width=.2\textwidth]{Figures/VLDLMstarEv.ps} }
\caption{Evolution as a function of the redshift of the twelve models with different total mass and  initial densities. The positions of the star particles are projected on the $XY$ plane (proper Mpc). The upper row corresponds to the high mass case, the central row to the intermediate mass case, and the bottom row to the low mass case. The initial density decreases from left to right. So the models displayed in the three groups of small panels in the left column are HDHM, HDIM, HDLM. The same in the other columns but for the ID-, LD- and VLD- cases, respectively. In each sub-panel the redshift changes from left to right and top to bottom: $z\simeq$ 11.5 (7 for VLDHM), 5, 3, 1.}\label{starev1}
\end{figure*}

\subsection{General remarks}\label{remarks}

Before going into the results of our study, we shortly discuss  here a few possible drawbacks of our models.

First, as already mentioned,  we adopt the stellar efficiency $\epsilon_{SF}=1$, in contrast to what observational data and simple theoretical considerations would suggest. This has proven to cause non-negligible effects on the dynamical evolution of the systems; they will be discussed in the  Sections below.

Second, no treatment of the multi-phase nature of the interstellar medium has been included. In contrast, different phases naturally develops within the gas (see Sect. \ref{dyn}). Furthermore,  it is long  known that numerical effects like  \textit{overcooling} may arise if the gas is modeled according to the standard SPH method \citep{Merlin2007}. Given these considerations,   a more detailed treatment of the interstellar medium not only is urgent but may also cure parte of the problems encountered with overcooling.

Third, we have not considered the  presence of nuclear activity, neglecting both the dynamical action of a central
super-massive black hole, and the energy feedback from an AGN, both of which are currently believed to play important roles in the evolution of a proto-galaxy.

Fourth, we have not included other sources of feedback apart from stellar - i.e., magnetic fields and their pressure or cosmic rays.

Finally, our cooling functions do not take into account the state of the Hydrogen gas; in particular, the variation of the molecular fraction of Hydrogen, which strongly influences the cooling efficiency below $10^4$ K in primordial, non-enriched gas (see Sect. \ref{cool}).

\section{Model results} \label{results}

All the simulations have been run on the HP Blade Server \textsc{Monster} with 14 hosts and 116 CPUs at the Astronomy Department of Padova. Each major galaxy model made use of four parallel CPUs, whereas companion test models of small size were run on single processors. On the whole, the simulations took $\sim 100,000$ CPU hours, requiring from 30,000 to 80,000 time-steps. There are two groups of models:

\textsf{First group: testing the models}. The first group  contains the test computations performed to assess the model response to varying \textit{some} important parameters. They are calculated with a smaller number of particles (usually from 5000 to 10000). To save computational time, not all the models have been calculated up to  the present age.  However all of them are carried out to the stage in which the mass assembly is completed (say at least a few Gyrs).
The following questions are addressed:

(i) \textit{Star formations efficiency}. These are the models already described in sect. \ref{starfo}.  They are calculated to check the sensitivity and overall behaviour at varying the SF dimensionless efficiency $\epsilon_{SF}$ of the star formation rate.

(ii) \textit{Stellar feedback}. A model is run with no  feedback from stars. As expected, the star formation rate grows dramatically at early times (see Fig. \ref{sfrNoF}; the run had to be stopped very soon because the continuous collapse of cold gas, not contrasted  by the action of heating due to feedback, requires smaller and smaller time-steps and therefore high computational costs). Our prescription for the stellar feedback can therefore be considered as reasonably efficient.

\textsf{Second Group: the reference models}. The  second group contains the reference set of models, with higher resolution. All of them are calculated well beyond the completion of the mass assembly, a few even down to the present age or very close to it. All important data referring to the models at the same reference redshift $z=1$ are summarized in Table \ref{tab2}, whereas in Table \ref{tab3} are displayed similar data but for the last computed stage of each model (different age and redshift).

\textsf{Results for the reference models}. In the following we shortly describe the main results for the twelve reference models. We start showing in Fig. \ref{starsb} a snapshot of their geometrical structure projected on the $XY$ plane at the last computed. In most cases the spatial distribution of stars resembles that of classical ETGs; in one case a filamentary substructure, and perhaps satellites companions are visible. Fig. \ref{sfr2} displays the star formation histories of the twelve models along their lifetime, down to the last simulated epoch. As found by \citet{Chiosi2002} and expected here, the kind of star formation at work changes with the galaxy mass and initial density. It is monolithic-like (a single dominant initial episode) in high mass and/or high initial density proto-galaxies, it gradually turns into a broad rather long and mild activity at decreasing mass/and/or initial
density, and finally it gets like a series of discontinuous very episodes (bursting mode) for low mass and/or low initial densities galaxies. The point will be developed in more detail below.

\begin{table*}
\caption{Properties of the twelve major models at $z=1$ (i.e. after
$\simeq 6$ Gyr of evolution). Left to right: total stellar mass,
total mass of the virialized halo, fraction of the total initial
mass composed by star, fraction of the initial gas converted into
stars, virial radius of the whole system (i.e. stars, Dark Matter
and gravitationally bound gas; it is the radius at which the mass
density is 200 times the background density at the epoch of
virialization, which we assume to have happened at $t \simeq 2$ Gyr
looking at the energy trends of the models: see Fig. \ref{energy}),
half mass radius of the stellar system projected on the XY plane,
axis ratios of the stellar system projected on the $XY$ plane.}
\centering \label{tab2}
\begin{tabular}{|l|l|l|l|l|l|l|l|}
\hline
Model & $M_*$ [$M_{\odot}$] & $M_{vir}$ [$M_{\odot}$] & $M_*/M_{vir}$ & $M_*/M_{gas,ini}$ & $r_{vir,tot}$ [kpc] & $r_{1/2,*}$ [kpc]& $b/a_{XY}$\\
\hline
\hline
HDHM & $7.8\times10^{11}$ & $1.1\times10^{13}$ & 0.071 & 0.27 & 153.0 & 15.7 & 0.53 \\
\hline
IDHM & $7.5\times10^{11}$ & $1.0\times10^{13}$ & 0.075 & 0.26 & 141.8 & 16.8 & 0.54 \\
\hline
LDHM & $7.4\times10^{11}$ & $9.4\times10^{12}$ & 0.080 & 0.26 & 133.8 & 14.5 & 0.57 \\
\hline
VLDHM & $6.3\times10^{11}$ & $7.5\times10^{12}$ & 0.080 & 0.22 & 112.5 & 11.2 & 0.52 \\
\hline
HDIM & $2.0\times10^{10}$ & $1.6\times10^{11}$ & 0.13 & 0.45 & 37.6 & 5.7 & 0.62 \\
\hline
IDIM & $1.9\times10^{10}$ & $1.5\times10^{11}$ & 0.13 & 0.43 & 35.7 & 5.8 & 0.63 \\
\hline
LDIM & $1.9\times10^{10}$ & $1.4\times10^{11}$ & 0.14 & 0.42 & 33.3 & 5.3 & 0.58 \\
\hline
VLDIM & $1.3\times10^{10}$ & $1.1\times10^{11}$ & 0.12 & 0.29 & 28.3 & 5.8 & 0.57 \\
\hline
HDLM & $1.2\times10^{8}$ & $2.6\times10^{9}$ & 0.028 & 0.17 & 9.2 & 2.2 &  0.69 \\
\hline
IDLM & $1.0\times10^{8}$ & $2.5\times10^{9}$ & 0.040 & 0.14 & 10.0 & 2.5 & 0.64 \\
\hline
LDLM & $8.9\times10^7$ & $2.3\times10^9$ & 0.021 & 0.13 & 11.8 & 2.2 & 0.56 \\
\hline
VLDLM & $5.0\times10^7$ & $1.7\times10^9$ & 0.011 & 0.07 & 10.5 & 2.7 & 0.58 \\
\hline
\end{tabular}
\caption{Properties of the twelve major models at the final stage of
the simulated evolution. Left to right: final redshift, final age,
total stellar mass, total mass of the virialized halo, fraction of
the total mass composed by star, fraction of the initial gas
converted into stars, virial radius of the whole system (i.e. stars,
Dark Matter and gravitationally bound gas; it is the radius at which
the mass density is 200 times the background density at the epoch of
virialization, which we assume to have happened at $t \simeq 2$ Gyr
looking at the energy trends of the models: see Fig. \ref{energy}),
half mass radius of the stellar system projected on the XY plane,
axis ratios of the stellar system projected on the $XY$ plane.}
\centering \label{tab3}
\begin{tabular}{|l|l|l|l|l|l|l|l|l|l|}
\hline
Model & $z_{end}$ & $t_{last}$ [Gyr] & $M_*$ [$M_{\odot}$] & $M_{vir}$ [$M_{\odot}$] & $M_*/M_{vir}$ & $M_*/M_{gas,ini}$ & $r_{vir,tot}$ [kpc] & $r_{1/2,*}$ [kpc]& $b/a_{XY}$ \\
\hline
\hline
HDHM & 0.22 & 11.0 & $7.5\times10^{11}$ & $1.5\times10^{13}$ & 0.050 & 0.26 & 153.0 & 15.6 & 0.56 \\
\hline
IDHM & 0.77 & 8.0 & $7.4\times10^{11}$ & $1.5\times10^{13}$ & 0.050 & 0.26 & 141.8 & 16.5 & 0.48 \\
\hline
LDHM & 0.50 & 8.7 & $7.3\times10^{11}$ & $1.5\times10^{13}$ & 0.049 & 0.25 & 133.8 & 15.8 & 0.57 \\
\hline
VLDHM & 0.83 & 6.6 & $6.3\times10^{11}$ & $1.3\times10^{13}$ & 0.048 & 0.22 & 112.5 & 11.2 & 0.52 \\
\hline
HDIM & 1.0 & 5.8 & $2.0\times10^{10}$ & $2.1\times10^{11}$ & 0.10 & 0.45 & 37.6 & 5.7 & 0.62 \\
\hline
IDIM & 0.75 & 7.0 & $1.9\times10^{10}$ & $2.1\times10^{11}$ & 0.08 & 0.43 & 35.7 & 5.8 & 0.63 \\
\hline
LDIM & 0.58 & 8.1 & $1.9\times10^{10}$ & $2.0\times10^{11}$ & 0.10 & 0.42 & 33.3 & 5.2 & 0.75 \\
\hline
VLDIM & 0.15 & 11.8 & $1.7\times10^{10}$ & $1.4\times10^{11}$ & 0.12 & 0.38 & 28.3 & 4.9 & 0.83 \\
\hline
HDLM & 0.36 & 9.7 & $1.5\times10^{8}$ & $3.3\times10^{9}$ & 0.045 & 0.19 & 9.2 & 2.3 & 0.74 \\
\hline
IDLM & 0.22 & 11.0 & $1.4\times10^{8}$ & $3.3\times10^{9}$ & 0.04 & 0.16 & 10.0 & 2.4 & 0.67 \\
\hline
LDLM & 0.05 & 13.0 & $1.4\times10^8$ & $3.2\times10^{9}$ & 0.04 & 0.19 & 11.8 & 2.1 & 0.79 \\
\hline
VLDLM & 0.0 & 13.7 & $1.0\times10^8$ & $3.0\times10^{9}$ & 0.03 & 0.10 & 10.5 & 2.7 & 0.65 \\
\hline
\end{tabular}
\end{table*}

\subsection{Dynamical evolution and final morphologies} \label{dyn}

All the models go through an initial phase of expansion, as they follow the Hubble flow. The central regions, in which the density peak is more pronounced, soon detach from the outward motion and start re-collapsing, forming the core of the virialized structure. As time goes on, more and more external regions turn around and fall onto the central virialized core. The process can be roughly described as a secondary infall, with time-scales strongly dependent
on the properties of the particular halo under consideration. More massive haloes globally virialize later with respect to smaller ones, as expected in the hierarchical cosmology. However, denser haloes also re-collapse sooner than the shallower ones, so the epoch of virialization is decided by the interplay between these two effects. At times depending on the depth of the central potential well, the central regions of the haloes begin to reach the critical density $\rho_*$ at which the gas becomes able to form stars. Due to the strongly peaked density perturbations, all models soon develop a central stellar system; however, in the very first stages of the evolution we note the formation of many small clumps of stars, which soon merge into a single entity. In most cases, additional smaller stellar systems form in regions far from the center because of local concentrations of cold gas, eventually merging onto the major body of the proto-galaxy. In model galaxies of low mass, sparse isolated stars form within clumps of expanding gas (wind), forming extended haloes of extremely low density around the central body of the proto-galaxy.

The whole history of the  building up of the stellar component of the model galaxies is shown in series of panels in  Fig. \ref{starev1}. Each panel show the projection on the XY plane (coordinates in proper Mpc) of the position of the star particles. Each row corresponds to a total mass (high, intermediate and low starting from) and each columns to an initial over-density: from high (left) to ver low (right). Finally each group contains four subpanels depending on the redshift at which the snaptshot are taken: from left to right and top to bottom the redshift is z=11.5 (7 for the very low density, high mass case), 5, 3, 1.

All models finally develop a central dense stellar system, located at the centre of the collapsed halo of DM. Some amount of spurious global bulk motion of the system (due to numerical instabilities) can be observed in some cases (see Figs. \ref{starsb} and \ref{starev1}). At $z=1$, most of the systems have almost completely relaxed into stable
configurations, although in the massive ones pronounced irregular features are clearly visible (arms of star forming regions, diffuse haloes of stars around the central object). These are \textit{not} formed by old populations infalling onto the central object, but by young SSPs escaping from the galaxy (this can be inferred by checking the birth age of the star particles and their velocities). The bright, spherical stellar clump visible in the bottom of the VLDHM model in Fig. \ref{starsb} has mass $M_*\simeq10^{10} M_{\odot}$, much larger than the masses of the low mass models. Such stellar clumps must therefore be considered tidal galaxies, formed by gas condensation. Many examples of such systems
are known to exist \citep[see e.g.][]{Elmegreen2007b}, even if they are generally thought to be generated within interacting systems. In our cases, local condensations of gas, compressed and pushed outwards by the pressure exerted by the inner regions of the system, collapse to form stars while being ejected from the galaxy. The tidal objects are finally engulfed by the massive galaxy, but usually leave a faint expanding trail of debris.

\begin{figure}
\centering
\includegraphics[width=.35\textwidth]{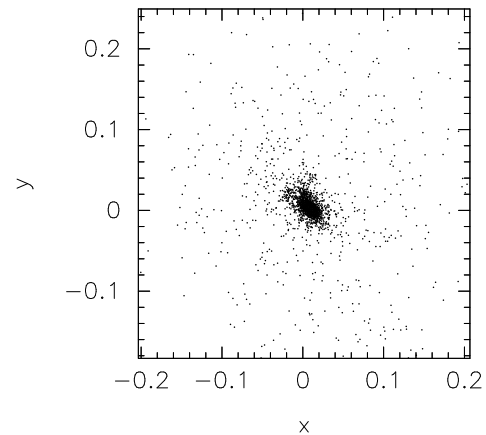}
\caption{Low mass models develop extended haloes of stellar particles around the central object (in this picture, model HDLM is shown at $z=1$; distances are in proper Mpc).}\label{diffusestars}
\end{figure}

On the other hand, low mass systems are surrounded by extended haloes of sparse stellar particles, as shown in  Fig. \ref{diffusestars} for model HDLM at $z=1$. Perhaps, these may be identified with the observed intra-cluster light
\citep[ICL,][]{Monaco2006}, which could therefore be the result of diffuse star formation within gas escaping from low mass galaxies, rather than the consequence of tidal stripping and/or disruption of bounded systems.

It is worth recalling here that a different choice for the dimensionless efficiency $\epsilon_{SF}$ may have lead to different dynamical histories, as discussed below. In particular, a lower efficiency could have reduced the formation of diffuse stellar haloes.

All galaxies have triaxial shapes; their axis ratio at $z=1$ and at the final redshifts, obtained from their projection on the $XY$ plane, are summarized in Tables \ref{tab2} and \ref{tab3}. If observed from this angle of view, all the models would be classified as E3 - E5 galaxies in the Hubble system. Obviously, more spherical shapes could be obtained changing the view angle, and almost all models could be morphologically classified as almost perfect spherical galaxies choosing a suitable projection. Noticeably, the minimal initial rigid rotation can hardly be considered the main cause for the pronounced triaxiality of the models, which must be therefore a consequence of their dynamical history. In
particular, observing the first stages of the formation of these systems it appears that their collapsing motions took place along the filamentary structures of the cosmological over-densities, and stars tend to conserve elongated orbits. This view is strengthened by the fact that reducing  $\epsilon_{SF}$, by requiring a higher degree of dissipation and condensation of gaseous particles, yields more concentrated and spherically-shaped structures. It would be worth investigating how this would affect to present day structure these systems. Perhaps, in this context, a non-negligible role could also be played by the absence of large scale tidal motions caused by the adopted prescription for the evolution of the whole proto-galaxy (the initial sphere) \textit{in void}.

The injection of energy from stars into the ISM causes turbulence and large scale motions in the gas (see for instance the situation illustrated in Fig. \ref{gasbubble}). Depending on the particular circumstances under which it takes
place, the stellar energy feedback can lead to the ejection of large amounts of gaseous material from the potential well of the galaxy, or to the formation of condensed structures, resulting in new star forming sites. The minor impact of feedback in low mass haloes ultimately results in the large differences in their star formation histories, as explained below.

\begin{figure}
\centering
\includegraphics[width=.4\textwidth]{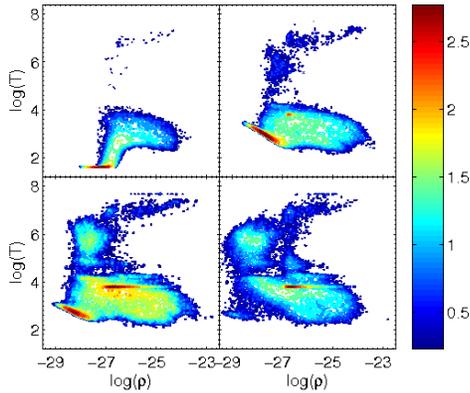}
\caption{Phase diagram density $\rho$ (g/cm$^3$) vs. temperature T (K) of gas particles for the IDIM model, at four stages of evolution ($z\simeq$10.0, 7.0, 5.0, 3.5). The hot gas decreasing its density is leaving the galaxy, due to stellar feedback.}\label{rhoTzvar}
\end{figure}

Different phases naturally develop in the gas during the first stages of the galaxy formation as clearly shown the phase diagram (temperature vs. density of the gas) displayed in  Fig. \ref{rhoTzvar} for the prototype model IDIM. First, the initially cold, expanding gas begins to heat to a typical temperature of T$\simeq10^4$ K in the central regions, as soon as they  detach from the Hubble flow and collapse. When the radiative cooling becomes effective, this gas begins to cool and to form stars. The ejecta from the SSP subsequently heat the surroundings\footnote{Once more,
it is worth pointing out that heating of the surrounding gas is due to the mechanical pressure exerted by the stellar particles, and \textit{not} to a redistribution of their thermal budget; see Sect. \ref{feedback}.}. Now, the infalling gas is forced to interact with these heated regions. If the hot gas is enough, this may lead to expanding shock fronts (see below). Thus, after several Myr of activity, the gas can be roughly split as follows. A cold and dense phase is formed by gas in the central regions of the galaxies, where stars are forming. On the opposite extreme, hot and thin gas \textit{either} is leaving the halo (galactic wind) after being heated by stellar feedback, \textit{or} it is infalling onto the central object, after being shock-heated because of the interaction with the central, hot regions. Two more phases are clearly detectable: the hot \textit{and} dense gas, where the stellar feedback is heating the gas at very high temperatures before these regions expand under the action of the local pressure, and the cold but thin gas belonging to the cosmological cosmic web (before re-ionization takes place, or cooling after re-ionization; see Sect.
\ref{reion}).

\begin{figure}
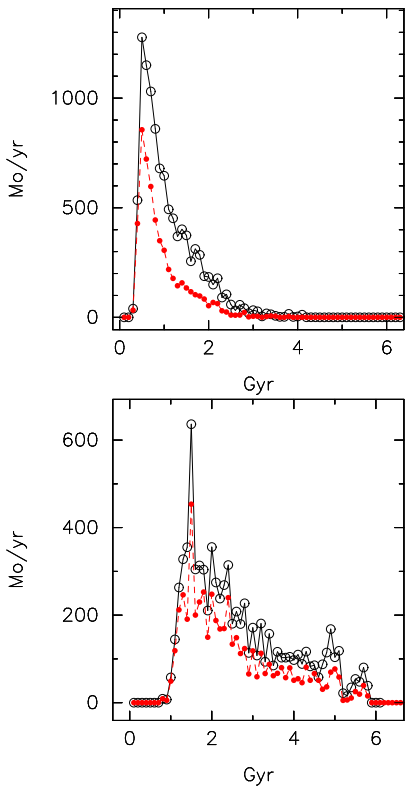

\centering
\includegraphics[width=.3\textwidth]{Figures/sfrHDHMinout.ps}
\includegraphics[width=.3\textwidth]{Figures/sfrVLDHMinout.ps}
\caption{Star formation histories of the whole galaxies (solid black line, large circles) compared with the corresponding ones within the central region ($r_{inner} \simeq 20$ kpc; red dashed lines, large points) for two large mass systems. Left panel: model HDHM.  Right panel: model VLDHM.}\label{sfrinout1}
\end{figure}

\begin{figure}
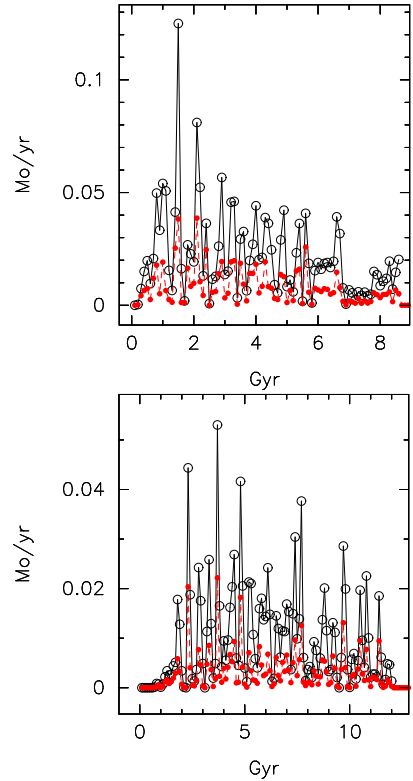

\centering
\includegraphics[width=.3\textwidth]{Figures/sfrHDLMinout.ps}
\includegraphics[width=.3\textwidth]{Figures/sfrVLDLMinout.ps}
\caption{Same as in Fig. \ref{sfrinout1}, but for two small systems (here $r_{inner} \simeq 2$ kpc). Top panel: model HDLM, bottom panel: model VLDLM.}\label{sfrinout2}
\end{figure}

\begin{figure}
\centering
\includegraphics[width=.45\textwidth,height=6cm]{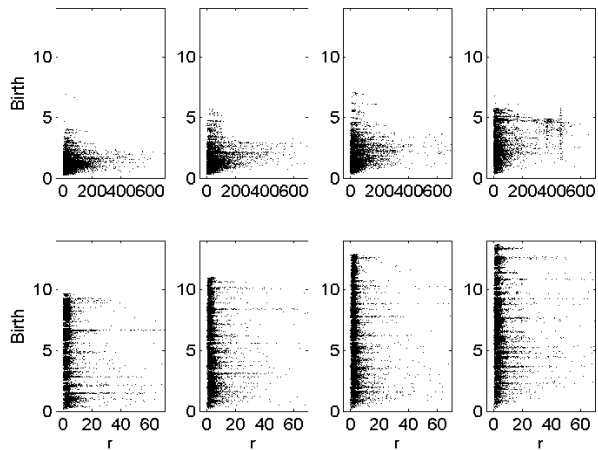}
\caption{Date of birth of the stars (in Gyr) as a function of their radial positions (distance from the barycenter in kpc) for the high and low mass models models (left to right: decreasing density; top to bottom: decreasing mass), at the final time-step of their evolution (the same age as in Fig. \ref{starsb}). The radial distance is calculated from the barycenter of the stellar system (kpc). The virial radii of the stellar systems are similar to the extension along the $X$ axis. }\label{agestar}
\end{figure}

\subsection{Star formation histories}

\begin{table*}
\caption{Properties of the star formation history for the twelve
models. Left to right: maximum peak of star formation rate [$M_{\odot}/$yr], epoch of
maximum activity [Gyr], percentage $p_*$ of the final stellar mass
assembled at four different redshifts, and epoch at which different
fractions $f$ of the final stellar mass (i.e. at $z=1$) has formed
[Gyr] (with corresponding redshift).} \centering
\begin{tabular}{|l|c|c|r|r|r|r|l|l|}
\hline
Model & SFR$_{max}$  & $t_{SFR_{max}}$ & $p_{*,z=10}$ & $p_{*,z=5}$ & $p_{*,z=2}$ & $p_{*,z=1.5}$ & $(t,~~z)_{f=1/2}$ & $(t,~~z)_{f=99/100}$\\
\hline
HDHM   & 1380   & 0.5 & 1  4.9 & 67.3 & 99.5 & 100.0 & 0.863 ~~6.5 & 3.029 ~~2.2 \\
\hline
IDHM   & 1020   & 0.6 &    2.9 & 55.2 & 98.0 & 100.0 & 1.097 ~~5.4 & 3.778 ~~1.8\\
\hline
LDHM   & 800    & 0.9 & $<$0.1 & 39.7 & 93.2 &  98.5 & 1.417 ~~4.4 & 4.398 ~~1.5\\
\hline
VLDHM  & 350    & 1.5 &    0.0 &  5.2 & 67.2 &  80.1 & 2.423 ~~2.7 & 5.083 ~~1.2 \\
\hline
HDIM   & 19.0   & 1.2 &    2.8 & 35.5 & 92.2 &  98.9 & 1.589 ~~4.0 & 4.392 ~~1.5 \\
\hline
IDIM   & 13.5   & 1.5 &    0.1 & 19.9 & 86.0 &  98.3 & 1.976 ~~3.3 & 4.530 ~~1.4 \\
\hline
LDIM   & 12.0   & 2.5 & $<$0.1 &  9.5 & 69.4 &  85.9 & 2.568 ~~2.6 & 5.806 ~~1.0 \\
\hline
VLDIM  & 8.0    & 2.5 &    0.0 &  1.1 & 42.8 &  66.3 & 3.537 ~~1.9 & 5.837 ~~1.0 \\
\hline
HDLM   & 0.135  & 1.5 &    2.3 & 17.9 & 61.5 &  76.1 & 2.780 ~~2.4 & 5.940 ~~1.0 \\
\hline
IDLM   & 0.075  & 3.0 &    0.6 & 10.2 & 49.1 &  69.6 & 3.480 ~~1.9 & 5.850 ~~1.0 \\
\hline
LDLM   & 0.06   & 2.0 &    0.2 &  6.2 & 54.9 &  79.2 & 3.000 ~~2.2 & 5.800 ~~1.0 \\
\hline
VLDLM  & 0.55   & 3.8 &    0.0 &  4.8 & 25.2 &  46.0 & 4.400 ~~1.5 & 5.800 ~~1.0 \\
\hline
\end{tabular}
\label{tab4}
\end{table*}

Looking at Fig. \ref{sfr2}, it is clear that the initial conditions of each halo deeply influence the star formation history (SFH) of the parent proto-galaxy. Even at a first glance, two clear trends can be easily identified.

First: \textit{the more massive the halo, the more peaked is the SFH}. All massive haloes $M_{tot}\simeq10^{13}M_{\odot}$) show a single, strong burst at very early times, followed by a slow decrease down to quiescence. A weak tail of activity persists down to later times in the less dense systems, thus accounting for a weak dependence on its density (or environment). Smaller haloes, in turn, display a more fragmented and prolonged
star formation activity, often alternated with periods of quiescence.

Second: \textit{lower densities correspond to delayed activity (the first burst starts later) and minor overall efficiency (the peaks of activity have lower magnitude)}. A delay up to $\simeq 1$ Gyr in the beginning of the SF activity can be seen in the low density models. Moreover, the maximum rate is generally halved, and the activity is
prolonged (this second effect is of crucial importance in intermediate mass haloes, $M_{tot}\simeq10^{11}M_{\odot}$).
Apparently, the mass of the halo is the dominant characteristic to determine its evolution, but intermediate mass haloes can have very different histories depending on their global over-density.

Table \ref{tab4} summarizes the properties of the SFHs. 
Interesting quantities are the intensity of star formation (in $M_{\odot}/$yr) at its peak value
and the time (in Gyr). Furthermore, the gradual building up of the stellar mass in a galaxy is measured by the percentage $p_*$ of the total stellar mass at redshift $z=1$ assembled at selected values of the redshift ($z$=10, 5, 2 and 1.5), and the time (redshift) at which 50\% and 99\% of the stellar mass at $z$=1 is assembled. This is indicated by the pairs $(t, z)_{f=1/2}$ and $(t,z)_{f=99/100}$ in Table \ref{tab4}. 

All the models begin forming Population II stars between $z\simeq 20$ and $z\simeq 7$. The HDHM halo experiences the strongest \textit{and} earliest peak of activity, reaching some $1.5\times10^3 M_{\odot}/$yr at $z \simeq 10$. More than 50\% of its total stellar mass is assembled \textit{before} $z\simeq6$. If observed now from
the Earth, this galaxy would appear massive ($M_* \simeq 3\times10^{11} M_{\odot}$) and essentially red at $z\simeq 4$. Its tail of SF activity continues down to $z\simeq1.6$, contributing 19\% of the total stellar mass. Looking at less dense haloes of the same mass, it can be noticed how the maximum of activity decreases in magnitude and is delayed to later times, down to a peak of $7.5\times10^2 M_{\odot}/$yr at $z\simeq 5$ in the VLDHM model, where moreover an intense activity is prolonged to much later epochs. Despite the similarity between the final objects formed from
massive haloes, a clear trend can be extrapolated from the data given in Table \ref{tab4}:
the lesser dense the halo, the later the mass assembly takes place (with the VLDHM halo assembles half of its total stellar mass at $z\simeq2.7$). The overall efficiency of the star formation process in these models is 20\%-30\%.

Intermediate mass systems apparently have more complex histories. Moving from denser to less dense systems, the shape of the star formation curve changes dramatically. Strongly over-dense haloes undergo an intense burst at early times, much as the most massive systems. However, less dense haloes gradually display more prolonged, and uneven histories, with many different peaks. Apparently, in this range of masses the initial density plays a dominant role in deciding the fate of the mass assembly of the galaxy. The efficiency of the star formation process is very high in these cases: 30\% -50\% of the initial gas is converted into stars.

All low mass systems show a continue, bursty SFH, sometimes with long  periods of quiescence. In these systems, the feedback from stars is sufficient to halt the star formation process soon after its beginning. However, this means that as soon as the first stars cease their strong feedback activity, the gas is able to re-collapse and form new generations of stars. This galactic ``breathing'' leads to systems in which many different generations of stars are present. The overall efficiency of the star formation mechanism is lower in this cases, with 10-20\% of the initial gaseous mass turned into stars (giving high $M/L$ ratios, comparable to those of many observed dwarf galaxies).

Figs. \ref{sfrinout1} and \ref{sfrinout2} compare the SFH of four models (HDHM, VLDHM, HDLM, VLDLM) \textit{across the whole systems} with the SFH \textit{within their central regions} (that is, within the central 20 kpc in massive systems and within the central 2 kpc in the small ones). It can be noticed that massive systems build their central regions before their outskirts, while in less massive galaxies the SF process is active both in the inner and in the outer regions during the whole lifetime of the galaxies. This is consistent with some recent observational findings
\citep{vanDokkum2010} on the way galaxies are built-up.

\begin{figure}
\centering
\includegraphics[width=.4\textwidth]{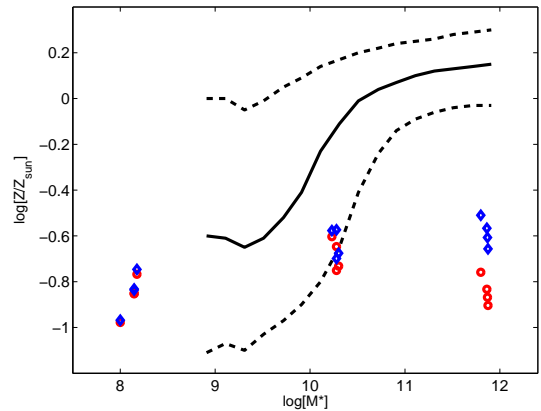}
\caption{Averaged metallicity $\log[Z / Z_{\odot}]$ versus stellar mass $\log M^*$ [$M_{\odot}$] for the twelve reference models, compared to the observed mass-metallicity relation. Black lines: data from \citet{Gallazzi2005} (solid line: median distribution of their sample; dashed lines: 16th and 84th percentiles). Red open circles: averaged metallicity of all stellar particles in each model. Blue open circles: averaged metallicity within the inner 5 kpc in each model.}\label{metals}
\end{figure}

\begin{figure}
\centering
\includegraphics[width=.4\textwidth]{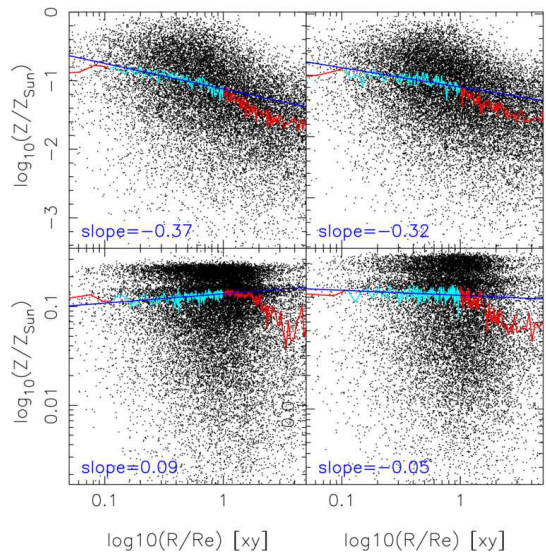}
\caption{Gradients in metallicity. Left to right, top to bottom: models HDHM, LDHM, HDLM, LDLM at their last computed snapshot. Plotted is the metallicity $\log[Z / Z_{\odot}]$ versus the fractional radial distance $r / r_{e}$ of all particles. Also shown are the median-binned metallicity profile, the number of particles being the same in each bin (blue and red solid lines, referring to the inner and outer region, respectively), and the linear best fit (black solid line; the slope is written in the left bottom corner of each panel). See text for details.}\label{grad_metals}
\end{figure}

\subsection{Stellar populations: ages, mean metallicities, and gradients in metal content} \label{populations}

\textsf{Stellar ages}. The distribution of the galactic ages ($T_{G,i}$) at which the star particles are created as a function of the radial distance is shown in Fig. \ref{agestar} limited to the high and low mass models (the behaviour of the intermediate mass ones falls in between these two). The real star particles ages are $T_*= T_G -T_{G,i}$. In early epochs, the stars are preferentially created in the central regions, then the star forming activity expands to larger radii (inside-out mechanism), and moving towards the present time, the stellar activity tends to shrinks again towards the centre. This simply mirrors the star formation history and the mechanism of mass assembly presented above.

\textsf{Mean metallicity}. Turning now to the metal content reached by the model galaxies, we look at the average stellar metallicity as a function of the total stellar mass. The results are shown in Fig. \ref{metals} where the red circles refer to all the star particles, whereas the blue ones only to those within the central 5 kpc of each model. The black lines are the median (solid) and the 16th and 84th percentiles (dashed) of the data presented in \citep{Gallazzi2005}. There is marginal agreement between data and theoretical results. The absolute values of the theoretical metallicities are too low compared with the observational estimates, in particular for the most massive models which are almost an order of magnitude too metal-poor. This may be due to several factors. First, the theoretical values are straightforwardly computed averaging the metallicity of all particles within the chosen radius without any further refinement; of course, observed metallicity are \textit{luminosity weighted}, while in our models the contributions from very metal-poor, primordial (old) and faint stars the same weight of more recent stellar populations. Second, part of the discrepancy may arise because of the different methods used to obtain observational and theoretical metallicities. For example, the use of the Kroupa IMF, which is tailored to fit the solar vicinity, may lead to underestimate the enrichment by massive stars in other environments. Recent studies seem to indicate that the IMF in massive ETGs is more skwed toward massive stars than commonly assumed \citep{Cappellari_etal_2012}  thus implying a net increase of metal production by massive stars in these systems. Third, the high star formation efficiency can again play a role, favouring the formation of low metallicity stars at early times and reducing the average metallicity of the models. Finally, as shown by \citet{Scodeggio2001} and discussed by \citet{Tremonti2004}, at least some part of the observed color- (and hence metallicity-) mass relation might just be a spurious consequence of an aperture effect, as observed colors (and metallicities) are usually measured within a fixed aperture, rather than a given fraction of the galaxy effective radius, so that the absolute values are not well determined. In conclusion, considering that chemical enrichment in NB-TSPH simulation is still far from being fully satisfactory, we are inclined to say that theory and data marginally agree each other.

In passing, we point out that the models naturally reproduce the observed trend of the $\alpha$-enhancement phenomenon in massive galaxies. The early peak of star formation in high mass systems, followed by quiescence, ensure the enhancement in light elements, whereas low mass systems with prolonged SFH are less enhanced, as it should be \citep[see the discussion in][]{Tantalo2002}.

\textsf{Metallicity gradients}.
Radial gradients in spectro-photometric properties of ETGs are known to exist \citep[see e.g.][ and references therein]{La_Barbera_etal2010}. The majority of ETGs in the local Universe features red cores \citep{La_Barbera_Carvalho_2009}, and less frequently blue cores, mostly found at low mass \citep{Suh_etal2010}. The radial gradients are interpreted as age and metallicity gradients of the constituent stellar populations \citep[e.g.][]{Gonzalez1993}.

In Fig. \ref{grad_metals} we plot the metallicity of each particle as a function of projected distance to the galaxy center (from left to right, and top to bottom, models LDHM, HDHM, LDLM and HDLM are displayed). For each model, the median-binned metallicity profile is shown, the number of particles being the same in each bin. In order to estimate the metallicity gradient, we performed an orthogonal least-square fit of the metallicity profile in the range $0.1 \,  r_{e}$ to $1  \, r_{e}$, i.e. the same range used in observational studies \citep{PVJ90}. The fitting was repeated for $100$ different two-dimensional projections, the average slope of the best-fitting lines giving the metallicity
gradient, $\nabla_Z = d\log Z/d\log  R$, of a given model. We repeated the computation  of $\nabla_Z$ by applying different weighting schemes, where each  particle was assigned no-weight, mass-weighted, and luminosity-weighted, finding no significant differences within the errors, especially at high mass (see Tab. \ref{tab:nablaz}).

Looking at the results in some detail, high-(low-) mass models exhibit metallicity gradients spanning the range from $-0.4$ ($-0.02$) to $-0.27$ ($0.16$), depending on the weighting scheme and environment. In general, low-density models tend to have more negative metallicity gradients. Moreover, at given density, high- (relative to low-) mass systems have more negative gradients.

In general,  the majority of the stars in the central regions of the massive systems have higher metallicity than those in the outskirts; on the contrary, in less massive systems the metallicity of the central stars is generally equal to that of the peripheral ones. Even if they are not straightforwardly visible in Fig. \ref{grad_metals}, interesting features to recall are: (i) on the average, the metallicity distribution of the star particles gets narrower at increasing radial distance; (ii) the maximum values are always reached in the centre, (iii) great deals of stars are made of recycled gas; (iv) finally the simulations often show the presence tidal satellites which have their own chemical history.

\begin{table*}
\caption{Metallicity gradients of four  ETG models in the radial range of  $0.1  \,  r_{e}$  to  $1  \,  r_{e}$.  Column  1  is  the  model ID label. Columns 2, 3, and 4, correspond to cases where the $\nabla_Z$ is computed with  no-weight, mass-weight, and luminosity-weight assigned to each particle. Errors are the rms of $\nabla_Z$ estimates among $100$ projections of each model.}
\begin{center}
\begin{tabular}{|c|c|c|c|}
\hline
 $ID$  &   $\nabla_Z$ (no-weight) &   $\nabla_Z$ (mass-weighted) &   $\nabla_Z$ (lum-weighted) \\
(1) & (2) &(3) & (4) \\
  \hline  \hline
LDHM & $-0.39 \pm 0.04$  & $-0.40 \pm 0.04$  & $-0.33 \pm 0.03$ \\
\hline
HDHM & $-0.32 \pm 0.04$  & $-0.31 \pm 0.04$  & $-0.27 \pm 0.03$ \\
\hline
LDLM & $ 0.00 \pm 0.04$  & $+0.01 \pm 0.02$  & $ 0.00 \pm 0.06$ \\
\hline
HDLM & $+0.09 \pm 0.02$  & $-0.02 \pm 0.01$  & $ 0.16 \pm 0.03$ \\
  \hline
\end{tabular}
\label{tab:nablaz}
\end{center}
\end{table*}

Although the metallicities of our models do not match observations in a absolute sense, the relative variations of metallicity (between, e.g., different masses, and/or different galacto-centric distances) are likely more robust, as they are less dependent on the absolute calibration of the ingredients the models rely on. This motivates for a comparison of the predicted $\nabla_Z$'s  with observational results. A steepening of the metallicity gradient with galaxy mass, qualitatively consistent with that seen in our models, has been reported  by \citet{La_Barbera_etal2010}, and, independent of galaxy environment, by \citet{La_Barbera_etal2011}. On the other hand, other observational studies, based on smaller samples, do not find a clear correlation of $\nabla_Z$ with mass \citep{Spolaor09}, or even a double-value behavior \citep{Tortora10}. For a galaxy mass of $3 \times10^{11}\,
M_\odot$ (similar to that of $\sim  7  \times10^{11}\, M_\odot$ of models HDHM and LDHM), La Barbera and collaborators found metallicity gradients of $-0.37 \pm 0.02$ and $-0.41  \pm  0.02$ for low- and high-density  ETGs, respectively. At low density, the estimate of is fully consistent with our model predictions (for no-weight and mass-weighted
$\nabla_Z$'s). At high-(relative to low-) density, the models exhibit shallower gradients, by $\sim 0.07$, in contrast to the data, where a (marginal) difference of $-0.04 \pm  0.03$ is found. For what concerns low-mass models, they exhibit almost null gradients, consistent with the observational finding of \citet{Spolaor09} that the $\nabla_Z$
vanishes at a galaxy mass of $\sim 10^9  \, M_{\odot}$. Whilst these results point to a fair agreement between our models and (some) available data-set, we postpone a more thorough investigation of both metallicity, age, and color gradients, as well as their correlation with other model properties, to a forthcoming contribution.

\subsection{Cold and hot gas accretion and ejection (galactic winds)} \label{accrete}

\begin{figure*}
\centering
\includegraphics[width=12cm,height=9cm]{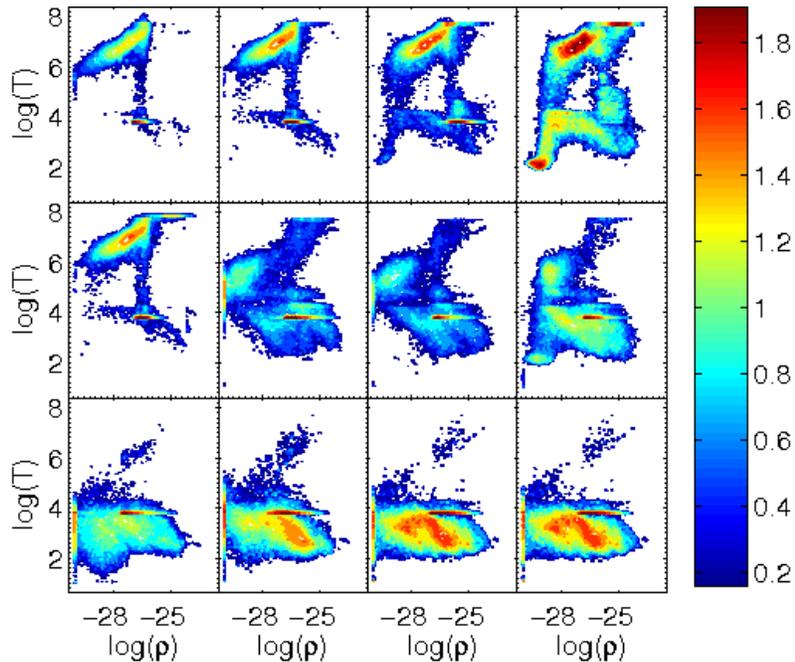}
\caption{Phase diagram density $\rho$ (g/cm$^3$) vs. temperature T (K) of gas particles for the twelve reference models, at redshift $z=2$. In each panel: left to right: high density, intermediate density, low density, very low density. Top row: high mass; intermediate row: intermediate mass; bottom row: low mass.}\label{rhoTz2}
\end{figure*}

Fig. \ref{rhoTz2} shows the gas  temperature vs. density diagram for models of our set at $z=2$. As expected
from theoretical considerations \citep{Rees1977}, a bimodal gas accretion can be observed in our models, as a function of the total mass of the proto-galactic halo. In low mass haloes, the gas accretes from the cosmic web without being shock heated to the virial temperature, and proceeds to flow along filaments towards the center of the halo, where it will eventually shock. In contrast, massive haloes soon shock the infalling gas to very high temperatures, suppressing the formation of local substructures and forming a central distribution of hot gas. haloes which do support shocks close to the virial radius are expected to contain a quasi-hydrostatic atmosphere of hot gas. \citet{Birnboim2003} claim that shocks can only form close to the virial radius in haloes with mass greater than $10^{11} M_{\odot}$ for primordial gas (or around $10^{12} M_{\odot}$ for gas of Solar metallicity), in good agreement with our models.

Looking at Fig. \ref{rhoTz2}, it can be noticed that the situation shows a strong dependence on the mass of the halo. In massive and dense systems, almost no cold gas (T$<10^4$ K) is present, so the star formation cannot proceed further. Moreover, an expanding shock front is evident (most of the gas is gathered in a narrow density strip moving outwards at high temperatures); this will result in a \textit{hot} galactic wind. The hot but thin gas is being shocked during its infall, in a hot accretion mode. The low density, cold tail is formed by gas expanding and never collapsing onto the
galaxy, cooling down after being photo-heated by reionizing radiation at $z=9$ (see Sect. \ref{reion}). More cold gas is present in less dense systems, both in high density regions (potentially forming new stars) and in the outskirts, the shock front being less pronounced. A large amount of dense ($\rho\simeq10^{-25}$ g/cm$^3$) and hot (T$>10^6$ K) gas is also present in massive systems which are still forming stars, due to the action of feedback.

In the intermediate mass systems, a large amount of cold and dense gas is still present, ensuring a continuing star formation. In low mass systems, almost no gas is heated to T$>10^5$. Thus, the gas is ejected from the galaxy at low temperatures, and no shock front develops.

\begin{figure}
\centering
\includegraphics[height=8cm,width=.4\textwidth]{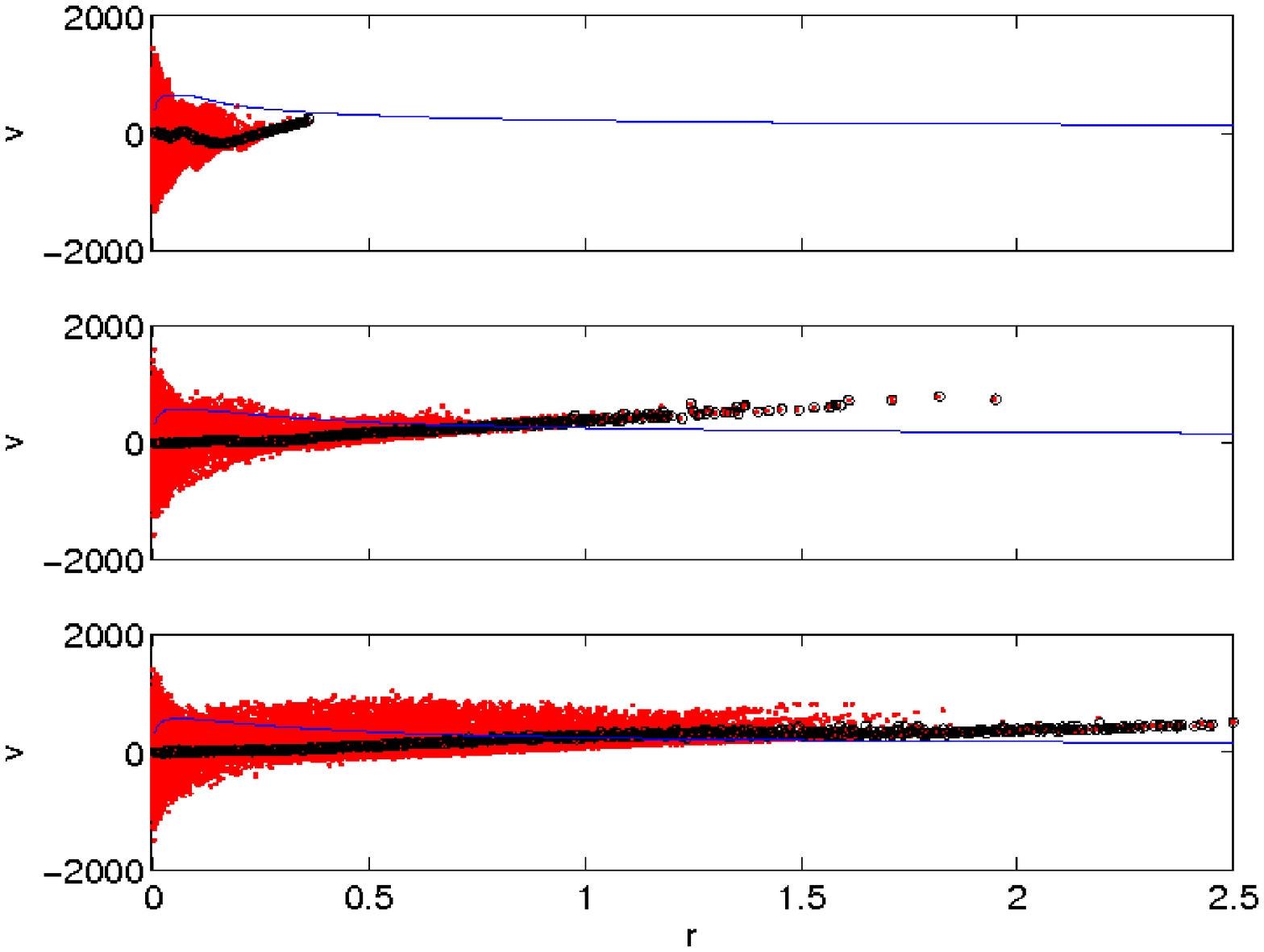}
 \caption{Galactic winds in the HDHM model. Red points: radial velocities of gaseous particles (km/s) against the distance from the barycenter of the systems (Mpc). Black circles: mean radial velocity in the spherical shell at radius $r$. Solid blue line: escape velocity as a function of the radius. Top to bottom: $z=4.4$, 1, 0.2.}\label{wind1}
\includegraphics[height=8cm,width=.4\textwidth]{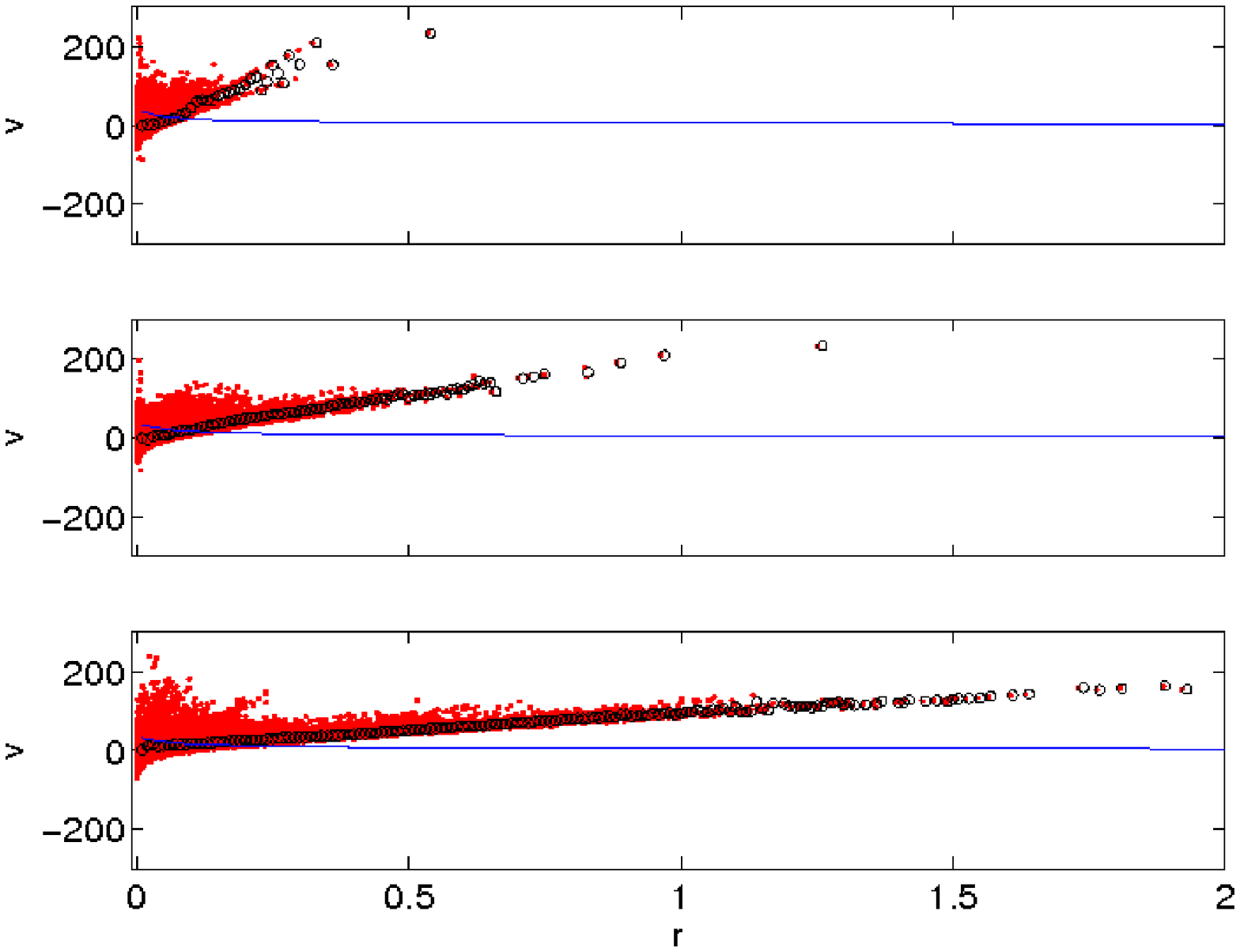}
\caption{Galactic winds in the LDLM model. Symbols as in Fig. 19. Top to bottom: $z=2.2$, 1, 0.05.} \label{wind3}
\end{figure}

\begin{figure}
\centering
\includegraphics[width=.45\textwidth]{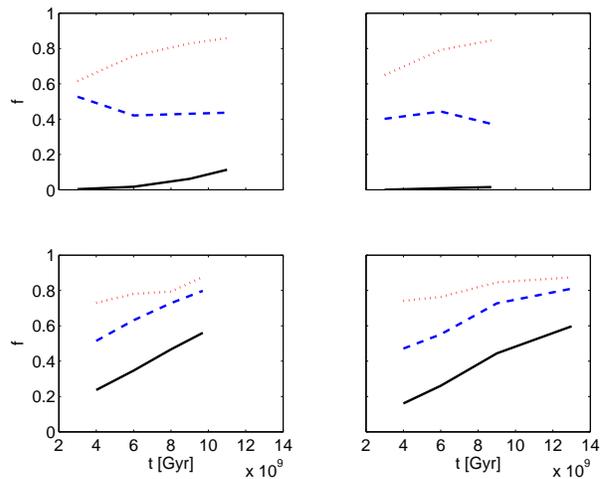}
\caption{Fractions of gas leaving the systems as a function of time. Top panels, left to right: HDHM, LDHM models. Bottom panels, left to right: HDLM, LDLM models. Red dotted lines: fractional mass of gas found outside $r_{ext}$ from the centers of the systems at the given time. Blue dashed lines: fraction of gas which has an outward-directed radial velocity, and is moving faster than the Hubble flow. Black solid lines: fraction of gas which also has radial velocity above the escape velocity of the system. See text for more details.}\label{fracwind}
\end{figure}

\begin{figure}
\centering
\includegraphics[width=.45\textwidth]{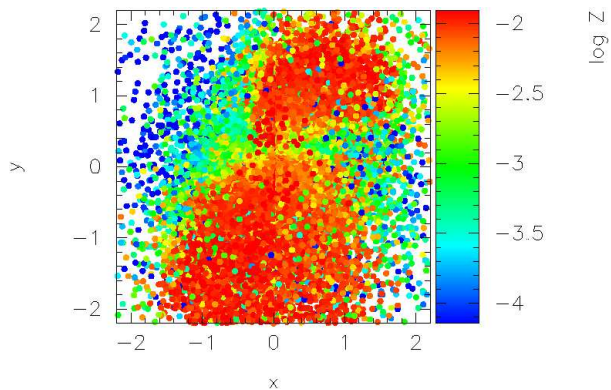}
\caption{Metallicity of the ejected gas (wind) in the HDHM model at $z_{fin}=0.22$. Distances are in proper Mpc. Chemically enriched gas (red particles), with $Z \simeq Z_{\odot} \simeq 0.019$, is leaving the galaxy (which
is located near the origin). Note the different metallicity, one or two orders of magnitudes lower, of the gas particles in the outskirts (green and blue): these do \textit{not} belong to the ``galactic wind'', since they haven't been ejected from the galaxy; they are part of the ``cosmic web''.}\label{windchem}
\end{figure}

\begin{figure}
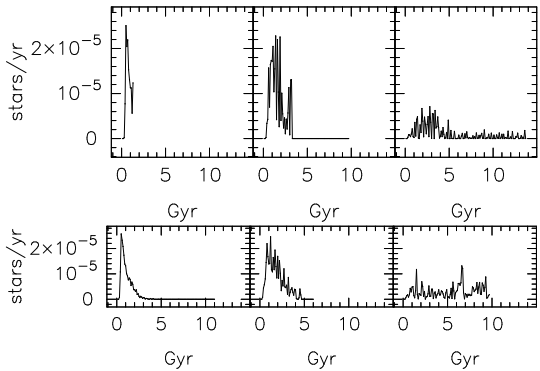

\centering
\includegraphics[width=.4\textwidth]{Figures/sfrLR.ps}
\includegraphics[width=.4\textwidth]{Figures/lrSFR2.ps}
\caption{Star formation histories for the three low resolution models, lrHDHM, lrHDIM, lrHDLM (in the bottom panels the corresponding SFH for the high resolution models are plotted for comparison). The rates have been normalized to the mass of the particles.}\label{sfrLR}
\end{figure}

\begin{figure}
\centering
\includegraphics[width=.45\textwidth]{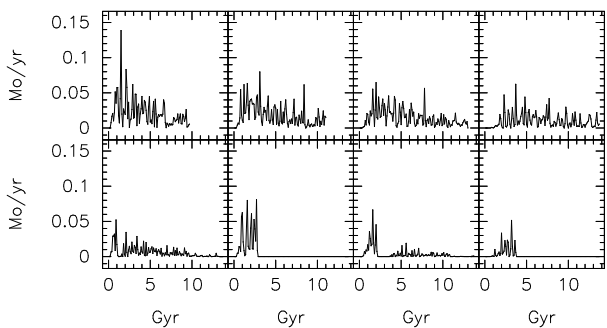}
\caption{Star formation histories for the low mass models, without (top panels) and with (bottom panels) imposing a minimum softening and smoothing length.} \label{sfr3giuste}
\end{figure}

\begin{figure}
\centering
\includegraphics[width=.4\textwidth]{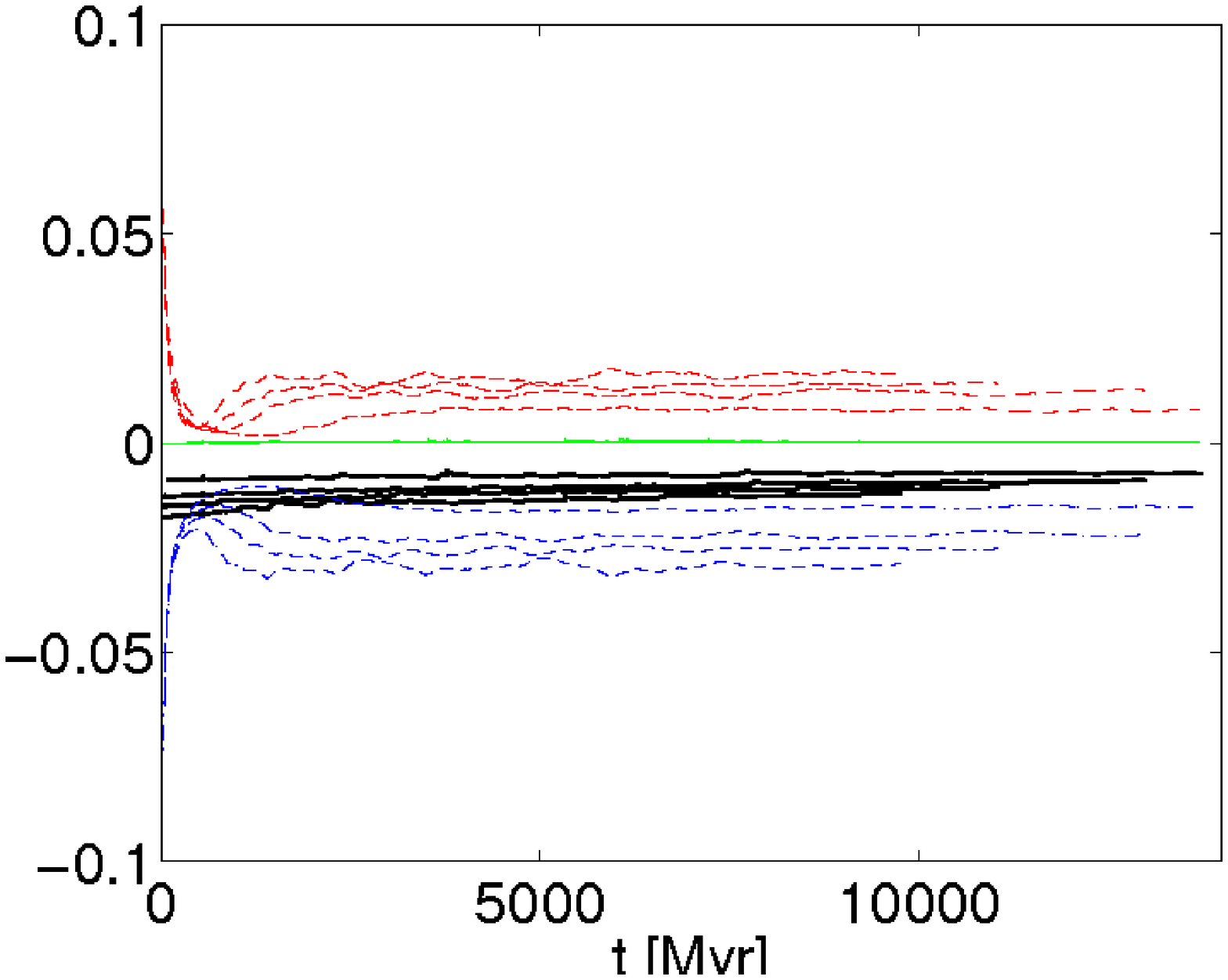},
\includegraphics[width=.4\textwidth]{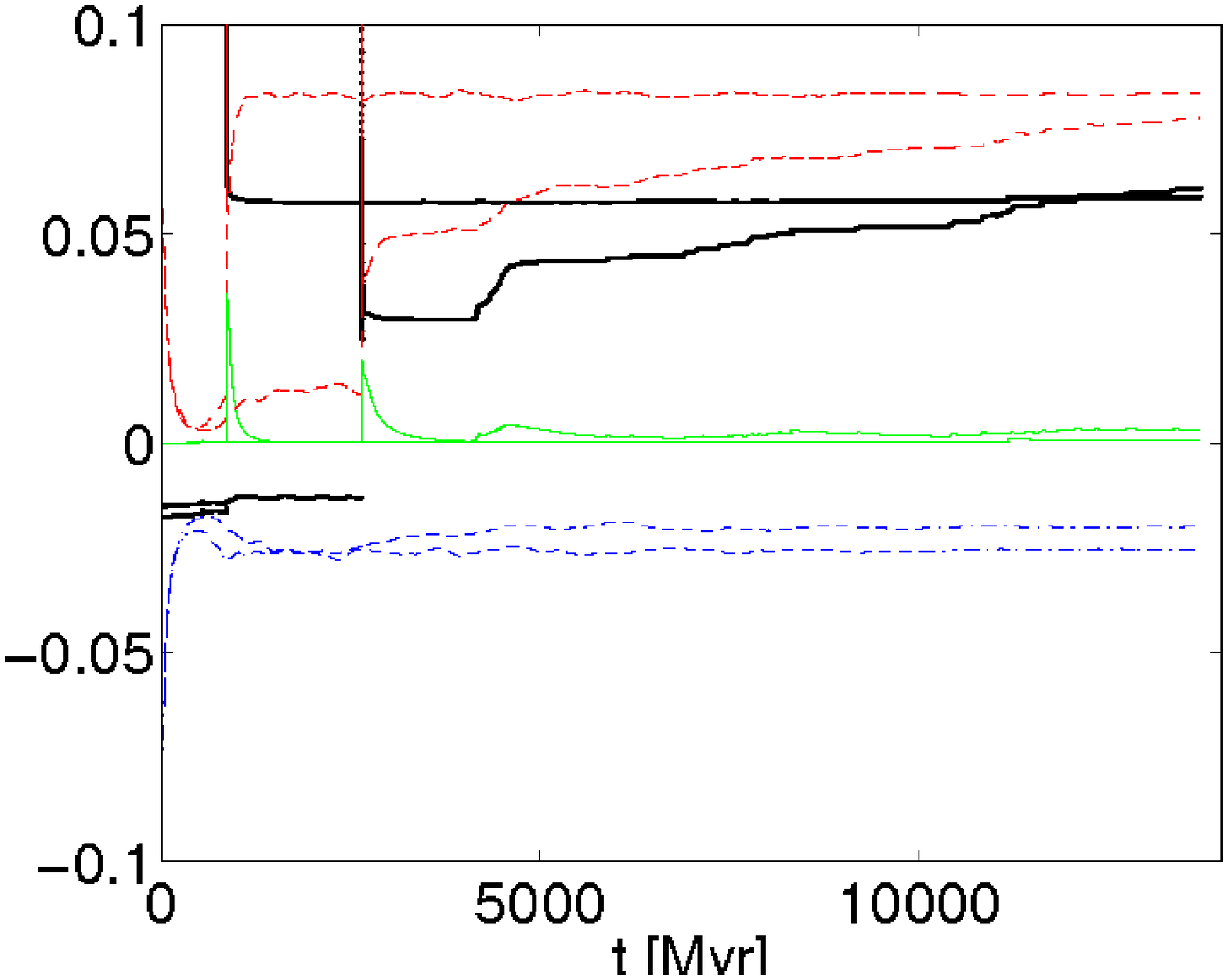}
\caption{Energy trends in the low mass models. In each panel the global values of the energies (in code units) are plotted versus time, superposing the results of different models. Left panel: models without minimum softening/smoothing lengths; right panel: models with minimum softetning/smoothing lengths ($\epsilon_{min}=h_{min}=100$ pc). Red dashed lines: kinetic energy; blue dashed lines: gravitational potential energy; green solid lines: thermal energy; black lines: total energy. } \label{energy}
\end{figure}

Figs. \ref{wind1} and \ref{wind3} plot the state of the gas in two reference models, namely HDHM and LDLM; they show the radial   velocity  as a function of the radial distance at three different epochs and compared to the escape velocity (the blue solid lines; see the caption for more details). The mass of the ejected gas increases with time in all the models. The average velocity is proportional to the mass of the system, with massive galaxies generating winds at $\simeq 2000$ km/s, and low mass galaxies at about one tenth of the speed. Interestingly, the average radial velocity is proportional to the distance, like in a Hubble-like flow; the ejection of the gas is therefore an explosive phenomenon, rather than a slow outflow. However, looking more carefully it can be noticed that there are different linear velocities at the same radial distances, meaning that many explosive events have taken place. There is also a huge mass of gas moving outwards at slower speed, at late times (particularly in the HDHM model).

In Fig. \ref{fracwind} we plot fractional amounts of escaping gas as a function of time, for four of the models (namely, HDHM, LDHM, HDLM, LDLM; the other models show similar behaviours). We  plot the trends \textit{only after} the systems have virialized; in other words, the total gravitational energy must have become constant to avoid the initial phases of galaxy assembly during which the gas accumulates toward the center and the galaxy is subject to episodes of quick expansions and/or contractions. The red dotted lines display the fractional mass of gas which is found \textit{outside} a given radius $r_{ext}$ from the centers of the systems at the given time. For massive systems we adopt $r_{ext} = 200$ kpc, while for small systems $r_{ext} = 10$ kpc (these values are similar to the virial radii of the models, see Tab. \ref{tab3}). The blue dashed lines are relative the fraction of gas which has an outward-directed radial velocity, and is moving faster than the Hubble flow at the correspondent redshift. Finally, the black solid
lines are the fraction of gas which is also above the escape velocity of the system. Clearly, two different regimes are at work. In the high mass realm, the fraction of gas which is \textit{really} leaving the galaxy is only a few percent. On the contrary, in low mass systems more than half of the total gas has sufficient velocity to escape. This is easily understandable. Assuming constant density in models of different total masses, the gravitational potential energy of a system is proportional to $M^{5/3}$, while the energy injection from SN explosions is proportional to the number of stars in the galaxy, i.e. roughly $\propto M$. Thus, the binding energy of a system grows more rapidly than the kinetic energy of its components, making it more difficult to escape from the system. However, this distinction is theoretical, and what we observe in the real Universe is simply the mass of gas that can be found outside a galaxy, moving outwards (i.e., the blue dashed line in the diagram). In this case, the situation in the two cases is much  similar,
however still with a higher efficiency in the low mass regime. The escaping gas is chemically enriched, as illustrated in Fig. \ref{windchem} for the HDHM model.

\subsection{Critical discussion of the results} \label{reliable}

Is the strong dependence of the SFH on the initial galaxy mass and density a physical result, or is it just a spurious numerical effect? The large difference in particle masses between the various simulations, a necessary choice to model galaxies spanning a wide range of sizes, is the major reason for this serious uncertainty. In high mass models, the gas particles have a mass $m_{HM} \simeq 5\times10^7 M_{\odot}$, whereas in the low mass models the mass of a gas particle is  $m_{LM} \simeq 1.2\times10^4 M_{\odot}$, i.e. a factor of $\simeq 4\times10^3$ times smaller. Does this difference
in mass resolution affect the global results of our study?

To check this issue, we run three low resolution models. The initial haloes for these test models were crudely obtained from the three high density models HDHM, HDIM, and HDLM, by simply considering one particle out four when reading the input files, and simultaneously multiplying the particle masses by a factor of four. Although this procedure is not rigorous, still it can be safely used to get a quick insight of the problem. The new models shortly indicated by lrHDHM, lrHDIM, lrHDLM have the total number of particles reduced by a factor of 4 and can be calculated on a single
CPU.  Their star formation is compared to that of their ``parent'' models and the results are plotted in Fig. \ref{sfrLR}. Even if some minor differences can be noticed, it is evident that the global trends are quite similar at the different resolutions. This is reassuring, because it implies that we can trust the results of our reference
models, at least as far as their different star formation histories is concerned. This also shows that meaningful results can be obtained even using a small number of particles with much less numerical effort.

Finally, we consider the effects of imposing  constant minimum threshold values to the softening and smoothing lengths of particles. To do so, we re-run the low mass models imposing minimum values $\epsilon_{min}=h_{min}=100$ pc. We plot the SFHs of the resulting models in Fig. \ref{sfr3giuste}, and compare them with the SFHs of the same models belong to the reference set in which no limitations have been imposed to the softening and smoothing lengths. Clearly, there are non-negligible differences. The global efficiency in the star formation process is lower in the models including the threshold. Moreover and perhaps even more important, the dynamical evolution is substantially different, because in
these models star formation is essentially halted after a strong peak of activity, whereas in the reference models it proceeds at nearly constant efficiency. To cast light on this issue we look at the temporal variation of the kinetic, gravitational potential, thermal, and total energies of the models under consideration. They are plotted in Fig. \ref{energy}. It is clear that imposing the limit values to the softening and smoothing lengths causes an uncorrect behaviour at the epoch of the strong ejection of hot gas following the first intense burst of star formation, when the dynamical stability of the system is put in troubles. The model response is not physically adequate to drive
the correct dynamical behaviour of the whole system. This means that the wrong choice of these parameters can deeply affect the overall evolution of the model. Variable softening and smoothing lengths are therefore highly
recommended. Nevertheless, a great deal of the general behavior of the reference models (in particular the gross features of the star formation history) still survives also in those with the constrained softening and smoothing lengths, thus securing that even with wrong choices for these two parameters we catch the essence of the problem.

\section{Mass density profiles} \label{profiles}

\begin{figure*}
\centering
\includegraphics[width=14cm,height=8cm]{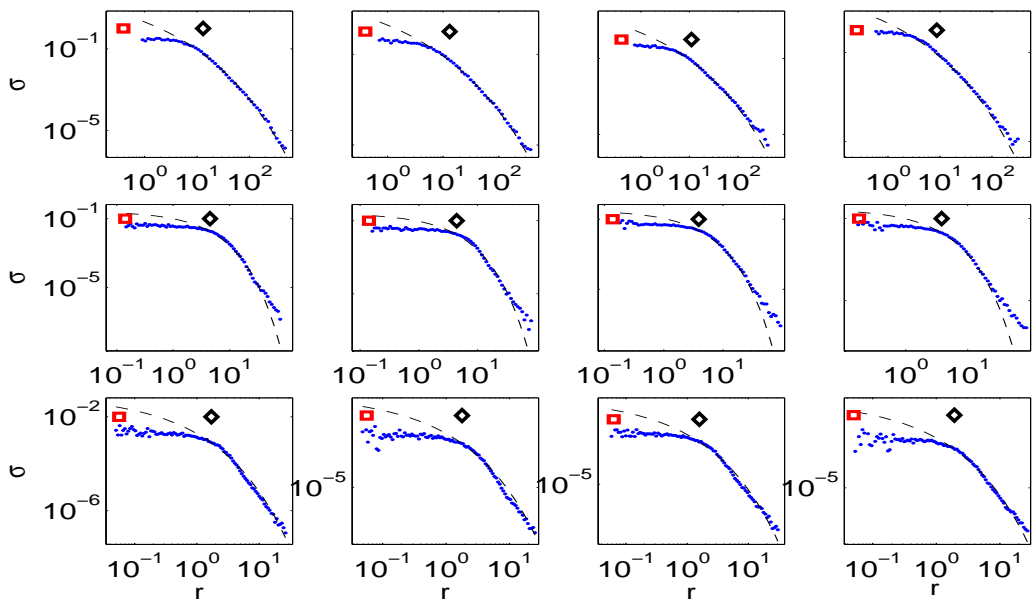}
\caption{Stellar surface density profiles for the twelve reference models at the final time-step of each simulation. Plotted is the projected surface density $\sigma$ [g/cm$^2$] against the logarithmic radius [kpc]. Blue dots: stellar particles radial averages; dashed line: best Sersic fit (see text for details). The black diamonds are placed at the effective radii $r_{e}$. The red squares are placed at the smallest value reached by adaptive softening lengths, thus
indicating the effective resolution of the simulations. Left to right: high density, intermediate density, low density, very low density. Top row: high mass; intermediate row: intermediate mass; bottom row: low mass.}\label{vaucsz1}
\end{figure*}

The geometrical structure of the model galaxies is best illustrated by the surface and volume density profiles. To compare these we choose a certain evolutionary stage, namely the model galaxies at a certain redshift, i.e. $z=1$. Fig. \ref{vaucsz1} displays the surface density radial profiles projected on the $XY$ plane at such redshift.
Over-plotted is the best \citet{Sersic1968} profile,

\begin{equation}
\sigma_S(r) = \sigma_0 \times e^{(0.324 - 2m)\left[\left(\frac{r}{r_{e}}\right)^{1/m} - 1\right]}
\end{equation}

\noindent where $r_{e}$ is the effective radius of the galaxy, as defined e.g. in \citet{Hernquist1990}, $\sigma_0$ is the surface density at $r_{e}$, and $m$ is the \textit{Sersic index} (if $m=4$ the de Vaucouleurs profile is obtained). All profiles are computed starting at $0.2\%$ of the virial radius of the galaxies to avoid the very central region of a galaxy where softening may introduce spurious numerical effects. The best-fitting Sersic index is $m \sim 4$, $m \sim 1.5$, and $m \sim 2.5$, for high-, intermediate-, and low-mass models, respectively. In other words, high-(relative to low- and intermediate-)mass models tend to have higher $m$, in qualitative agreement with the existence of a
luminosity-Sersic index relation of ETGs \citep{Caon1993}. However, one should notice that the most massive ellipticals in the local Universe tend to have $m \sim 8 $ \citep[e.g.][]{Ferrarese2006b}, while we find $m \sim 4$. Moreover, our intermediate-mass models have somewhat lower $m$ than  the low-mass ones, which is only marginally consistent with
luminosity-$m$ relation, considering its large scatter.

The overall agreement between the models and the Sersic curves is good in the external regions. However, a clear departure from the expected fits is evident in all models at small radii (some fraction of the effective radius, indicated with the black diamond shown in each panel). The central regions of the model galaxies tend to
\textit{flatten out} with a plateau in the mass density profile. Given the adaptiveness of the force softening, this feature can hardly be ascribed to numerical artifacts. The close scrutiny of the problem, suggests us  ascribe it to the high value of the dimensionless efficiency, $\epsilon_{SF}=1$, that we have adopted for the star formation rate.

\begin{figure*}
\centering
\includegraphics[width=14cm,height=8cm]{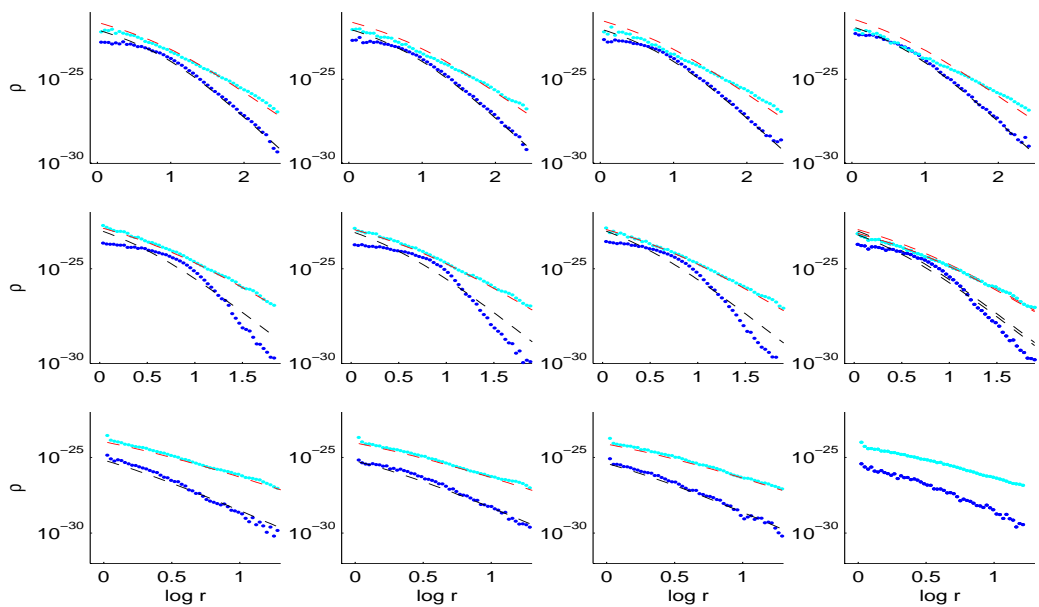}
\caption{Density profiles for the twelve reference models at $z\simeq1$. Plotted is the density $\rho$ [g/cm$^3$] against the logarithmic radius [kpc]; blue dots: stellar particles; cyan dots: DM particles; dashed lines: best Hernquist and NFW fits. Left to right: high density, intermediate density, low density, very low density. Top
row: high mass; intermediate row: intermediate mass; bottom row: low mass.}\label{densz1}
\end{figure*}

This conclusion is strengthened by the three-dimensional mass density profiles  shown in Fig. \ref{densz1}. These profiles are computed starting at $0.05\%$ of the virial radii, and they are compared with the best fits of the Hernquist \citep{Hernquist1990} and \citep{Navarro_etal_1996} profiles (NFW), for stars and DM, respectively. While DM haloes smoothly follow the expected NFW profiles, the density of the stellar component never exceeds that of the DM, and the central regions of the model galaxies are dominated by the DM, contrary to the expectations \citep[e.g. ][ but see also Grillo, 2010]{Tortora2009, Padmanabhan2004}.

The relations between the efficiency of star formation and the gas density at which star form is straightforward. If star formation proceeds slowly (low values of $\epsilon_{SF}$, the gas can get high densities before it gets consumed by star formation itself. The opposite if star formation proceeds at high efficiency, high values of $\epsilon_{SF}$. As a consequence there is an immediate correlation between star formation and the dynamical behavior of a galaxy. Allowing gas particles to turn into stars as soon as they fulfill the required numerical criteria (that is, with unit efficiency) will generally result in quite diffuse stellar systems, with shallow potential wells and large effective radii. On the contrary, a low star formation efficiency leads to dense clumps of cold gas which give birth to more compact stellar clouds (and to a delay in the beginning of the activity).

Interestingly, this puts a constraint on the modality in which gas collapses and forms stellar systems. Historically, there are two possible and competing theoretical scenarios to describe the assembly of the stellar mass of an elliptical galaxy (excluding the possibility of merger events). In the first one, stars originally form far away from the effective radius of the galaxy and subsequently accrete onto its inner regions, in a non-dissipative fashion. In the other case, first the gas flows into the central regions and then it is turned into stars. If the correct
scenario is the first one, smoothing of the central density cusp of the profiles is expected, because the infalling stellar clumps loose orbital energy through dynamical friction and  the halo heats up \citep{Bertin2003}. In the second case, the central distribution is expected to steepen up, because the gas radiates away its internal energy and an adiabatic contraction takes place \citep{Gnedin2004}. Obviously, both processes are expected to play a role in reality, and this is indeed what we see in our models, in which however the final density profiles show a plateau in the central regions, as expected in the ``infalling stars'' model.

Our results also clearly suggest that if stars are formed too early, the resulting stellar systems are too shallow, ending up in ``wrong'' profiles: stars will tend to conserve their velocity dispersion as they fall into the potential wells and will therefore have elongated orbits, which is not the case when gas is let dissipate part of its kinetic energy and reach very high densities before turning into stars.

In this context, it is worth recalling that there is some observational evidence for a deficit in the central luminosity of a few early type galaxies as recently found in observational data by \citet[][their Fig. 1]{Cote2007}, and references therein). However, the observed flattening begins at smaller radii (a few percent of $r_{e}$), and it is present only in bright (i.e. massive) systems, whereas the small systems seem to present a luminosity excess. Indeed, looking  at the panels in the bottom row of Fig. \ref{vaucsz1}, one can notice that the central profiles of
low mass models show some scatter and tend to steepen in the very central regions (especially in the VLDLM model). Recalling that the observational information stands on the luminosity profile, given the presence of young and hence luminous stellar populations in the cores of our low mass galaxy models, (and their absence in the cores of the massive ones), we can speculate that the trends of the observed curves should be fairly reproduced by translating the mass
profiles into luminosity into luminosity profiles. This is an interesting point, because the  excess or deficit of light could be explained just in terms of the stars and DM orbits, without the presence of a super-massive black hole affecting the global dynamics in the center of the systems, as instead suggested in the studies quoted above.

\begin{figure} 
\centering
\includegraphics[width=8cm,height=8cm]{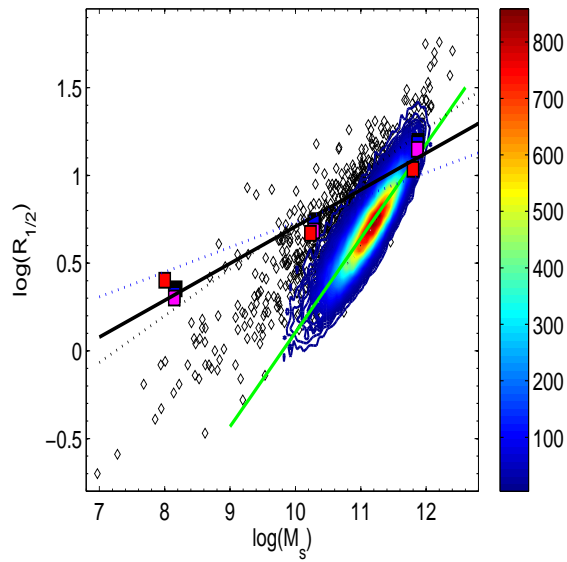}
\caption{Mass-radius relation, i.e. the relation $\log(r_{1/2})$ versus $\log M_{12}$, where $r_{1/2}$ is the half mass (light) radius, and $\log M_{12}$ is the total stellar mass in units of $10^{12} M_{\odot}$. The contour lines display the regions of the MR-plane populated by the same number of objects (the bar yields the correspondence between color-code and number of galaxies). The data are from the HB sample of \citet{Bernardi2010}. The green solid line is their linear best fit, $\log r_{1/2} = 0.54 \log M_{s} - 5.25$. The black  diamonds are the data from \citet{Burstein1997}, which also include the dwarf galaxies. Large squares: the twelve major models at the last simulated age; the same colors refer to same initial densities. Black: high density. Blue: intermediate density. Magenta: low density. Red: very low density. The thin dashed black lines are fits to the models (see text).} \label{reffmass}
\end{figure}

\section{The Mass-Radius Relation} \label{MRR}

To conclude our analysis, we briefly consider the Mass-Radius Relation (MRR) for ETGs. An overview of recent results concerning estimates of the star mass and dimensions of ETGs is given by the MRR shown in Fig. \ref{reffmass} which displays the sample of ETGs collected by \citet[][green dots]{Bernardi2010}. The sample contains about 65,000 objects. We have also added the small group of galaxies (about 600 objects in total) studied long ago by \citet{Burstein1997} to include also some dwarf galaxies. Owing to the huge difference in the total populations of the two sample cno cross checking of the data has been applied for the galaxies in common. This plot is only meant to provide an idea of position of real galaxies onto the MR-plane.  We also plot the position of the reference models coding them with different colors according to the initial over-density.

At a first glance, one would say that high and intermediate mass models fairly agree with the observations, whereas the low mass ones apparently have too large radii with respect to their masses. Therefore if on one hand the dimensions of intermediate and high mass models agree with the data, thus improving upon the situation with the hierarchical scheme that predicts too small radii, on the other hand because of deviation shown by the low mass models, the slope of the observational MRR cannot be accounted for. However, the thorough analysis of this issue made by \citet{Chiosi_etal_2012} has shown that the observational MRR is the result of a  subtle interaction among different concurring physical causes.

We begin by noting that as expected models with equal initial over-densities fall along the  straight lines $\log r_{1/2} = \alpha  \log M_s + \beta$, with slope $\alpha$ ranging from 0.2 to 0.3 and $\beta$ depending on the redshift $\beta \propto (1+z)^{-1}$. Slopes close to 0.33  are indeed what is predicted for collapsing spherical perturbations of DM \citep[see e.g. ][]{Fan2010}. The deviation from the pure $\log r_{1/2} = \alpha log M_s + \beta$ relationship is larger towards galaxies of lower mass and it is due to the complex baryon physics; the discussion by \citet{Chiosi2002} and \citet{Chiosi_etal_2012} on the issue still applies also here.

\citet{Chiosi_etal_2012} argue that the observed MRR for ETGs is the result of two complementary mechanisms. On one hand, the mass function of DM haloes hosting the visible galaxies gives (i) the typical cut-off mass at which, at any redshift, haloes become ``common'' on a chosen spatial scale, and (ii) the typical epoch at which low mass haloes begin to vanish at a rate higher than that at which they are born because of merger events \citet{Lukic2007}. On the other hand, these constraints define two loci (curves) on the MR-plane, because to each mass and redshift a typical dimension (i.e., radius) can be associated (using the basic relation between  mass and radius  of a collapsing object). If the typical dimension of a galaxy is somehow related to that of the hosting DM halo (as our NB-TSPH models seem to suggest), then the region of the MR-plane between the two limits fixed by the halo mass function is populated by galaxies whose dimensions are fixed at the epoch of formation, and only those objects that are ``possible'' at any given epoch may exist, populating a narrow region of the MR-plane. All details of this rather complicate analysis can be found in \citet{Chiosi_etal_2012} to whom the reader should refer.

\section{Summary and conclusions} \label{conclusions}

We have presented a numerical investigation on the star formation histories of ETGs, performed with the aid of
hydrodynamical NB-TSPH simulations of galaxy formation in cosmological context. By varying the initial properties of a single model in terms of total mass and initial over-density, we investigated how the initial configuration influences the formation of the system along the Hubble time.

The main result of our study is that the star formation history of a galaxy is directly dependent on its initial total mass and initial over-density. Massive galaxies ($M_{tot}\simeq 10^{13} M_{\odot}$) build their stellar content passing through an initial burst of activity at early times. In contrast, low mass systems ($M_{tot}\simeq 10^{9} M_{\odot}$) continue to form stars throughout their lifetime, with episodes of intense activity separated by period of relative
quiescence. Moreover, galaxies with the same initial mass have different star formation histories depending on their initial over-density: the deeper the perturbation, the more peaked, early and intense is the star forming activity. This is particularly true for  intermediate mass galaxies ($M_{tot}\simeq 10^{11} M_{\odot}$). Thus, denser objects, which may be identified with cluster and group central galaxies, have more peaked and earlier burst of star formation,
while less dense objects (i.e., field and satellite galaxies) usually have delayed and lower activity. This is largely consistent with previous analyses by \citet{Chiosi2002}, \citet{Clemens2006}, and \citet{Thomas2005} to mention a few.

The structural properties of our models are in good agreement with those of the observed real galaxies, with some important differences (e.g. the density profiles in the central regions, the mass-to-metallicity relation) which we ascribe to the unusual choice we have made for the star formation dimensionless efficiency parameter, $\epsilon_{SF}=1$.

The picture emerging from this study is that an early-hierarchical, monolithic-like intense activity of star formation at high redshift is possible starting from physically sounded  cosmological conditions, and is fully adequate to explain the observed features of massive elliptical galaxies.

Major mergers are certainly possible, but they are not strictly required. It is indeed easy to understand that on the contrary they would alter the subtle game among different physical ingredients all contributing to shape the  body of properties and their their mutual tight correlations shown by early type galaxies such as for instance the MRR itself \citep{Chiosi2002, Chiosi2012}. Along this line of thought \citet{Buonomo2000} showed how a major merger would shift the position of the parent objects towards the upper region of the MR plane, where no observed galaxy is found. Moreover, chemical features of the elliptical galaxies cannot be reproduced if (wet) merging events are considered. Finally, the shape of the galaxy mass function \citep{Bundy2006,PerezGonzalez2008} rules out the possibility that late major dry mergers are responsible for the mass assembly of massive systems, out of smaller but already passively
evolving objects. Our models show instead that massive galaxies can form out of single, peaked perturbations, consistently with the cosmological background.

Another interesting point to note is that  the stellar feedback seems to be sufficient to quench the star formation
process in massive objects \textit{without} invoking more exotic sources of energy, like Active Nuclei, provided it
is modeled with sufficient accuracy. Since the models are evolved \textit{in void}, the lack of late-time infall of gas could explain why the presence of an AGN is not strictly necessary. In relation to this, one could  argue  that strong galactic winds moving outward at high velocity are probably sufficient to inhibit  flows of material towards the inner regions. However, the occurrence of Active Nuclei cannot be excluded.

Furthermore, we are also able to recover the complex, episodic star formation histories typical of many dwarf galaxies. The galactic breathing phenomenon is  a  natural consequence of an accurate treatment of the stellar feedback process.

We conclude that the fate of a proto-galaxy is essentially determined by its initial conditions in terms of mass and over-density, while external factors such as encounters, mergers and disruptions, while substantially altering the evolution of the involved systems, are not a fundamental ingredient in the global evolution of the early type galaxies population.

\textbf{Acknowledgements}
The authors are grateful to the anonymous referee who greatly helped to improve upon the first version of the manuscript.
This study has been financed by the
University of Padua under the Strategic Research Project ``Algorithms and Architectures for Computational Science and Engineering''.


\bibliographystyle{mn2e}           

\bibliography{mnemonic,Galassie_biblio_Rev}    

\label{lastpage}

\end{document}